\documentclass[a4paper,11pt]{article}
\pdfoutput=1 

\usepackage{jheppub} 

\usepackage[T1]{fontenc} 

\usepackage{tikz}
\usepackage{tabularray}

\newcommand{\SU}{\mathrm{SU}}

\newcommand{\ket}[1]{\left|{#1}\right\rangle}
\newcommand{\braket}[1]{\left\langle{#1}\right\rangle}

\newcommand{\fp}{F^{+}}

\newcommand{\tr}{\mathrm{Tr}}
\newcommand{\cN}{{\cal N}}
\newcommand{\nFour}{$\cN \, = \, 4$ }
\newcommand{\rar}{\, \rightarrow \,}
\newcommand{\cO}{{\cal O}}
\newcommand{\eqsp}{\, = \,}
\newcommand{\la}{\langle}
\newcommand{\ra}{\rangle}
\newcommand{\cC}{{\cal C}}
\newcommand{\beq}{\begin{equation}}
\newcommand{\eeq}{\end{equation}}

\title{Higher-rank sectors in the hexagon formalism  and marginal deformations}

\author[a]{Burkhard Eden,}
\author[a]{Dennis le Plat,}
\author[b]{Anne Spiering}

\affiliation[a]{Institut f\"ur Mathematik und Physik, Humboldt-Universit\"at zu Berlin,\\
Zum gro{\ss}en Windkanal 6, 12489 Berlin, Germany}
\affiliation[b]{Niels Bohr Institute, University of Copenhagen,\\ 
Blegdamsvej 17, 2100 Copenhagen Ø, Denmark}

\emailAdd{eden@math.hu-berlin.de}
\emailAdd{diplat@physik.hu-berlin.de}
\emailAdd{anne.spiering@nbi.ku.dk}

\abstract{The hexagon approach provides an integrability framework for the computation of structure constants in \nFour super Yang--Mills theory in four dimensions. Three-point functions are cut into two hexagonal patches, on which the excitations of the long-range Bethe ansatz of the spectrum problem scatter. To this end, the Bethe states representing the operators also need to be cut into two parts. 
In rank-one sectors such entangled states are fairly straightforward to construct so that most applications of the method have so far been restricted to this simplest case. 
In this article we construct entangled states for operators in $psu(1,1|2)$ sectors, importing a minimum of information from the nested Bethe ansatz. The idea is successfully tested against free field theory for a sample set of correlators with up to three higher-rank operators. 

Further, we take a look at the same correlators in the presence of marginal deformations of the theory. 
While a systematic modification of the hexagon procedure remains out of reach for now, in practical applications the undeformed amplitudes are surprisingly efficient, especially for a certain one-parameter deformation.
}

\preprint{HU-EP-22/40}

\begin{document} 
\bibliographystyle{JHEP}
\maketitle
\flushbottom

\section{Introduction}

The AdS/CFT conjecture \cite{1,2,3} states that the maximally supersymmetric non-Abelian gauge theory in four dimensions  ($\mathcal{N}=4$ SYM theory) is equivalent to IIB string theory on an $AdS_5 \times S^5$ background. Since the field theory is useful as a weak-coupling description of the system, while the string theory captures strong coupling, how can the duality be put to the test? The BMN construction \cite{bmn} presented the first quantitative check in that a set of composite operators in the field theory was identified as the dual of a certain class of string-theory states. The leading quantum correction to the conformal dimensions in field theory agrees with the energy levels in string theory, although one has to bear in mind that the effective coupling is small in one case and large in the other. 

In the field-theory picture, the quantum corrections to the scaling dimension of gauge-invariant composite operators are discussed. In the \nFour model the elementary fields transform in the adjoint representation so that they are (matrix) products of fields under a gauge-group trace. To match with string results, the \emph{length} of these operators (i.e.\ the number of fields) has to be large, with most of them being scalar fields of the same type (\emph{vacua}) and only a few \emph{impurities}, i.e.\ other fields in between. In the simplest case, the impurities are all scalars of the same type. One obtains an operator-mixing problem whose eigenoperators have well-defined scaling dimension. The spectrum of planar one-loop anomalous dimensions in such an $su(2)$ \emph{sector} is equivalent to the energy eigenvalues of the \emph{Heisenberg spin chain} \cite{niklasDila,Minaza}. Bethe famously solved this eigenvalue problem by Fourier transform: every \emph{excitation} (impurity) is given a \emph{rapidity}, and when one of them overtakes another a phase factor arises that is commonly termed the $S$-\emph{matrix}. 

In the integrable model for the full spectrum problem of planar \nFour SYM theory, excitations of sixteen flavours (four scalars, four derivatives, four left fermions, and four right fermions) scatter on a chain of vacua. The so-called \emph{rank-one sectors}, like the $su(2)$ sector, have only one type of excitation, and here the spectrum of eigenstates can rather straightforwardly be obtained from the Bethe ansatz.
The $su(2)$-sector Bethe ansatz has been extended to the full set of composite operators of the planar \nFour model and arbitrary loop order \cite{beiStau1,beiStau2}, at least for a so-called asymptotic regime.
There are two approaches: first, all flavours are treated democratically and one writes a $16 \times 16$ component $S$-matrix for their scattering. Specifically, after having selected a field acting as a vacuum site, say a scalar $Z$, let us assume that there are $n$ flavours of fields moving on this chain. The scattering of two magnons $\chi^A(u_1)$ and $\chi^B(u_2)$, where $A,B \in \, \{1 \ldots n\}$ label their flavour and $u_{1,2}$ their rapidities, is described by the $S$-matrix elements $S^{AB}_{CD}(u_1,u_2)$ and results in a sum of two-magnon states with different flavour combinations, i.e.\ $S^{AB}_{CD}(u_1,u_2)\, \ket{\chi^C(u_2) ,\chi^D(u_1)}$. Thus the $S$-matrix becomes a true matrix of phase factors, as a flavour change is possible during the scattering. In addition, one writes a \emph{multi-component wave function} with one component for every initial ordering of flavours. Every component comes with a coefficient which will eventually be determined by a set of matrix Bethe equations alongside the Bethe roots. 
The second procedure is the \emph{nested Bethe ansatz}. Here, in order to diagonalise the problem one analyses the system layer by layer sorting by the types of excitations, at the cost of introducing \emph{auxiliary rapidities}. 
One identifies a sequence of raising operators in the symmetry algebra: the first will send the vacuum to a \emph{level-1} excitation, then by a second raising operation to a \emph{level-2} excitation, and so on. Then the level-2 magnons move on the sub-chain of level-1 excitations and not directly on the original chain of vacua.
This yields a system of coupled Bethe--Yang equations from which the spectrum can more easily be inferred than from the matrix problem.

Three-point functions became accessible by integrability methods through the hexagon approach introduced in \cite{BKV}.
In this approach the three-point function (or three-string vertex) is cut through the middle of the plane of the drawing, cf.\ Figure 1, yielding two hexagonal patches. The black edges are string pieces, or in the dual field-theory picture one half of a spin chain. Between these \emph{physical edges} lie \emph{virtual edges} which are coloured in the sketch. In the free field theory they can be understood as bunches of configuration space propagators stretching between the three operators. The number of propagators forming such an edge is dubbed \emph{edge width}. These quantum numbers play an important r\^ole most of all in loop computations. Yet, we will also encounter them in this work in the context of free field theory structure constants.
\begin{center}
\includegraphics[height = 2.5 cm]{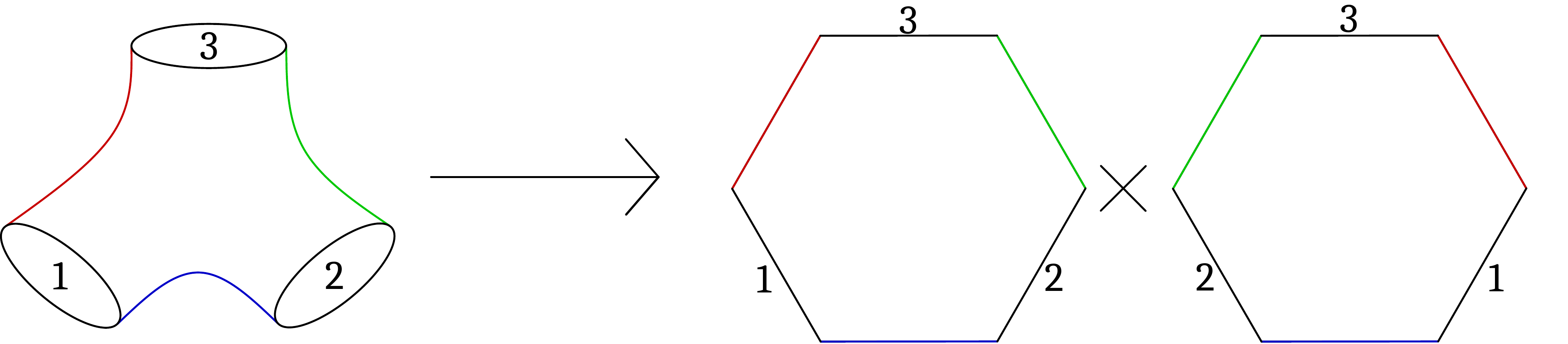}
\vskip 0.1 cm
\textbf{Figure 1}: Splitting a three-point function into two hexagons.
\end{center}
\vskip 0.2 cm
Considering closed string theory as a two-dimensional field theory, the authors of the earlier work \cite{romualdZoli} formulated axioms this three-vertex ought to obey. In \cite{BKV} similar axioms were posed for the two halves of the vertex, the hexagons in the sketch above. Here it was possible to find an explicit solution whose correctness has by now been supported by a good number of tests \cite{testHexagon1,testHexagon2}.

Having singled out the solutions of the Bethe equations \cite{Minaza,beiStau1,beiStau2} characterising the operators at the three outer points, the hexagon approach \cite{BKV} provides a recipe for computing the corresponding structure constant. First, excitations are distributed over the two parts of the respective physical edge. This is called preparing an \emph{entangled state} \cite{tailoring}.
One then looks at the sum over all these \emph{partitions} on the hexagon pairs, each weighted by a \textit{splitting factor}. How to evaluate each hexagon is explained in the original work \cite{BKV}, the main steps are as follows: First, all excitations are moved to the same edge by \emph{crossing transformations}. Then the excitations or \emph{magnons} are split into a left and a right part, based on methods developed in \cite{beiStau1,beiStau2}. 
Third, in the spectrum problem one would now use the $S$-matrix of \cite{beisertSu22} to scatter all the left parts and all the right parts amongst themselves. However, here scattering takes place only on one of the chains (no matter which), with the same $S$-matrix but an adapted overall normalisation. Finally, the left and right parts thereby obtained are contracted, and these momentum-depending contributions are summed over all the partitions.  
The resulting expression should be equal to the structure constant as computed by Feynman graphs, up to standard normalisations.

The salient property of the hexagon approach is that structure constants --- like energies in the Bethe
 ansatz --- can be extracted from the \emph{Bethe roots} alone, i.e.\ the rapidities of the excitations along the spin chain. By contrast, in any control calculation by field-theory means, the explicit form of the eigenoperators is needed. The \emph{Bethe wave function} provides such states when put on shell, so when evaluated on a solution of the Bethe equations, and are eigenstates of the \emph{dilatation operator} of the theory \cite{niklasDila,christophDila}. 
 
As far as we are aware of the literature, all but one of the applications of this method are concerned with \emph{rank-one sectors}. This includes the aforementioned $su(2)$ sector, the $su(1|1)$ sector with one type of fermions as excitations, and the non-compact $sl(2)$ sector with Yang--Mills covariant derivatives in an irreducible spin representation. 
In order to evaluate structure constants in higher-rank sectors, the nested hexagon approach \cite{nestedHexagon} was introduced, where the \emph{hexagon operator} -- especially in the presentation of appendix F in that paper -- is to some extent replaced by a wave function of the nested Bethe ansatz. In this work we instead aim at using the original hexagon approach with higher-rank operators at all three external points, while maintaining the matrix picture for the hexagon operator. More specifically, we will construct these external states as multi-component wave functions whose explicit form we fix via the nested Bethe ansatz.
For the evaluation using hexagons this will merely result in larger sets of partitions whose splitting factors will now consist not only of powers of momenta and the entries of the $S$-matrix, but also the information imported from the nested Bethe ansatz. 

In the second half of this article we ask whether the hexagon formalism can accommodate for the $\gamma$-\emph{marginal deformation} of the \nFour theory introduced in \cite{Frolov:2005dj}.\footnote{Historically, all exactly marginal $\cN=1$  deformations, including the  $\beta$-deformation, were classified for \nFour SYM in \cite{beta}. The $\beta$-deformation was later understood as a $TsT$- transformation \cite{Lunin:2005jy} on the string-theory side and then further generalised to the $\gamma$-deformation in \cite{Frolov:2005dj}.} For the real $\beta$-deformation, the nested Bethe ansatz of the spectrum problem can be deformed by decorating some of its various $S$-matrices (rather phases) by factors $e^{i \, \beta}$, and can further be generalised to the non-supersymmetric $\gamma$-deformations \cite{raduNiklas}.
Nevertheless, twisting the symmetries seems to clash with the hexagon construction in two ways: first, the $S$-matrix on the left or the right chain alone ceases to satisfy an axiom called \emph{physical unitarity}, i.e.\ the fact that scattering the same two neighbouring particles twice should reproduce the original state.\footnote{In the spectrum problem this is reconciled by the product structure. We are grateful to C.~Ahn for a discussion about this issue.} Second, to arrange the operators in the three-point function at the positions $0,1,\infty$ along a line in Minkowski and internal space, the \emph{twisted translation} \cite{BKV}
\beq
{\cal T} \eqsp -i \, \epsilon_{\alpha \dot \alpha} P^{\dot \alpha \alpha} + \epsilon_{\dot{a} a} \, \mathfrak{R}^{a \dot{a}} \label{eq:TwistedTranslation}
\eeq
with translation operator $P$ and internal-space rotation operator $\mathfrak R$, was introduced in the undeformed model.\footnote{The symmetry algebra of the three-point problem is a diagonal subalgebra given by those generators that commute with ${\cal T}$.} It mixes the spin-chain vacuum $Z$ with a second set of \emph{longitudinal} scalars $Y,\bar{Y}$ yielding a \emph{co-moving vacuum}
\beq
\hat Z(a) \, = \, Z + a \, (Y-\bar Y) + a^2 \bar Z \, , \label{twistedT1}
\eeq
where $a$ is the position of the field along the line. 
The inhomogeneous superposition of operators it creates at point $a=1$ is what is needed in free field theory to match with the hexagon results.
Up to now we have not understood how to consistently deform this transformation, and in particular dealing with operators containing the longitudinal scalar fields $Y, \bar Y$ is most subtle.
For a large part of the article we therefore study operators in $psu(1,1|2)$ sectors which contain one of the \emph{transversal scalars} $X, \bar{X}$, i.e.\ those which cannot be contracted on the co-moving vacuum \eqref{twistedT1}, and two types of fermions as excitations. In faint analogy to what was done in \cite{raduNiklas} we experiment with undeformed hexagons decorated by deformation factors keeping the \nFour twisted translation.

\vspace{0.5cm}
\noindent
The article is organised as follows:
\begin{itemize}
\item In the next section we define the sectors and correlators that we wish to study. We explain in detail how to import information from the nested Bethe ansatz into the multi-component wave functions of the $S$-\emph{matrix} picture, and how to construct the entangled states in higher-rank sectors.
\item In Section \ref{chapter3} we evaluate structure constants in \nFour super Yang-Mills theory using our hybrid formalism, i.e.\ combining the multi-component wave function and the nested Bethe ansatz.
\item In Section \ref{secMarginal} we review marginal deformations of the theory, focusing on the real $\gamma$-deformation, and
\item in Section \ref{chapter4} we look for consistent deformations of the previously discussed three-point functions.
\item We finally dare to take a glimpse at longitudinal scalars as well as loop corrections in the deformed setup in Section \ref{secSu2}.
\end{itemize}

\section{States in \nFour SYM theory} \label{chapter2}

\subsection{Conventions and definitions}
\nFour SYM theory contains six complex scalars $\Phi^{AB}=-\Phi^{BA}$, with R-symmetry indices $A,B\in\{1,2,3,4\}$. It will be useful to introduce the notation
\beq
Z \, = \, \Phi^{34}\,, \qquad X \, = \, \Phi^{24} \, , \qquad Y \, = \, \Phi^{14}\, ,
\eeq
where bosons $Z$ will later be identified as spin-chain vacuum states, and $X$ as a transversal and $Y$ as a longitudinal excitation. Moreover, there are four fermions $\Psi^A$ and four anti-fermions $\bar\Psi_A$, as well as the covariant derivative $D^\mu$ with space-time index $\mu\in\{1,2,3,4\}$. The four $Q$ supersymmetries that leave the vacuum $Z$ invariant are
\beq
Q_\mathrm{magnon} \, \in \, \{Q_1, Q_2, \bar Q^3, \bar Q^4 \}
\eeq
and transform the scalars into fermions as 
\beq
\delta Q_2 \, X \, \equiv \delta Q_2 \, \Phi^{24} \, = \, \epsilon^2 \, \Psi^4
\eeq
etc.\ ($\bar Q$ varies $\bar X$ in the same way), so that
\beq
X \rar \Psi^4, \, \bar \Psi_1 \, , \qquad \bar X \rar \Psi^3, \, \bar \Psi_2 
\eeq
under the magnon group. Note that $Y$ transforms into the same fermions under $Q_\mathrm{magnon}$. 

The largest closed subsector of the \nFour model without length-changing interactions enjoys a $ psu(1,1|2)$ symmetry. The first sector we consider is spanned by the fields $\{X, \Psi^4, \bar{\Psi}_1,D^{2\dot{2}}\}$ and its algebra contains the $\SU(2)$ group with R-charge lowering operator $\mathfrak{R}_3^2$ and the two supersymmetries $Q_2, \, \bar Q^3$. 
Using the tensor product notation of \cite{beiStau1,beiStau2}, we can represent the fields as bi-fundamental representations of $ su(2|2)^2$
\begin{align}
X \rar \phi^2 \otimes  \phi^{\dot{2}} \, , && \Psi^4 \rar \psi^2 \otimes  \phi^{\dot{2}} \, ,&& \bar \Psi_1 \rar \phi^2 \otimes \psi^{\dot 2} \, , &&  D^{2 \dot{2}} \rar \psi^2 \otimes \psi^{\dot{2}} \label{secX}\,,
\end{align}
where $\phi^a, \psi^\alpha$ is a fundamental representation of the first copy of $ su(2|2)$ and $\phi^{\dot{a}}, \psi^{\dot{\alpha}}$ of the second. Note that the subalgebra should only act on one of two spinor components of each fermion $\Psi^{4\alpha}$ and $\bar\Psi_1^{\dot{\alpha}}$; the actual pick will only alter non-vanishing results by global signs and for definiteness we choose $\alpha=2$ and $\dot{\alpha}=\dot{2}$ in \eqref{secX}. Furthermore note that we have dualised the flavour index of $\bar \Psi_1$ to an upper 2 by $\varepsilon^{21}$. Likewise, our second sector contains
\begin{align}
&\bar X \rar \phi^1 \otimes \phi^{\dot{1}} \, , && \Psi^3 \rar \psi^1 \otimes  \phi^{\dot{1}} \,, && \bar \Psi_2 \rar \phi^1 \otimes \psi^{\dot 1} \,, && D^{1\dot{1}} \rar \psi^1 \otimes \psi^{\dot 1}~. \label{secBar}
\end{align}
Finally, in the two sectors \eqref{secX}, \eqref{secBar} --- and indeed the entire super spin-chain model \cite{beiStau1,beiStau2} --- four of the two-component fermions appear, but not their conjugates.
The construction for longitudinal scalars $Y$ and $\bar{Y}$ is analogue.

We will evaluate three-point functions with operators inserted along the line $(0,0,a,0)$ at the positions $a=0,1,\infty$. In order to do so, we construct all operators at $a=0$ and then translate them with the help of the twisted-translation operator \eqref{eq:TwistedTranslation} which contains the lowering operators $\mathfrak{R}_3^1, \, \mathfrak{R}_4^2$ of the internal flavour symmetry $ su(4)$. It acts on $Z$ as \eqref{twistedT1}, whereas for the transversal and longitudinal scalars one finds
\begin{align}
&\hat{X}(a)=X~, &&\hat{\bar{X}}(a)=\bar{X}~,&&\hat{Y}(a)=Y+a\, \bar{Z}~, &&\hat{\bar{Y}}(a)=\bar{Y}-a\, \bar{Z}~.
\end{align}
The resulting \emph{effective} propagator for the co-moving vacuum is $\la \hat Z \hat Z \ra \eqsp 1$, and there exist non-vanishing off-diagonal propagators $\la \hat Z \hat Y \ra, \la \hat Z \hat {\bar Y} \ra$. 
The possibility to contract two operators with fermions is similarly provided by the \emph{twisted translation} \eqref{eq:TwistedTranslation} which acts as
\begin{align}
&\hat \Psi^4(a) = \Psi^4 + a\,  \Psi^2 \, , &&\hat {\bar \Psi}_2(a)  = \bar \Psi_2 - a\, \bar \Psi_4 \, , &&\hat \Psi^3(a)  = \Psi^3 + a \, \Psi^1 \, , &&\hat {\bar \Psi}_1(a)  = \bar \Psi_1 - a\,  \bar \Psi_3 \, .
\end{align}
Hence the effective fermion propagators like $\langle \hat{\Psi}^4(a_1) \hat {\bar \Psi}_2(a_2)\rangle$ are proportional to $1/(a_1-a_2)^2$, as is the bosonic propagator $\langle \hat X(a_1) \hat{\bar{X}}(a_2) \rangle$. We will have to rely on their existence also when trying to deform the model. This is certainly a stumbling block, but hopefully not a fatal one as the fermions are \emph{transversal excitations} \cite{BKV} like $X, \bar X$, and the specific longitudinal operator configurations that we study, will project onto the transversal case.

In the following, we will consider three-point functions of higher-rank operators of the form
\begin{eqnarray}
&& \cO_A^L \eqsp \tr(\hat Z^{L-2} X \hat \Psi^4) \, , \qquad \cO_{\tilde A}^L \eqsp \tr(\hat Z^{L-2} \bar X \hat {\bar \Psi}_2) \, , \nonumber \\
&& \cO_B^L \eqsp \tr(\hat Z^{L-2} \hat \Psi^4 \hat {\bar \Psi}_1) + \tr(\hat Z^{L-2} X \, D \hat Z) + 
\tr(\hat Z^{L-1}  D X) \, , \label{oABC} \\[1 mm]
&& \cO_{\tilde B}^L \eqsp \tr(\hat Z^{L-2} \hat \Psi^3 \hat {\bar \Psi}_2) + \tr(\hat Z^{L-2} \bar X \, D \hat Z) +  \tr(\hat Z^{L-1}  D \bar X) \, , \nonumber \\[1 mm]
&& \cO_C^L \eqsp \tr(\hat Z^{L-4} X X \hat \Psi^4 \hat \Psi^4) \, . \nonumber
\end{eqnarray}
Here we leave the dependence on $a$ implicit. Note that these expressions only sketch the field content in the operator mixing -- any actual operator is a linear combination of many terms with relative positions of the magnons different from the cases displayed above and non-trivial coefficients. We will also need the rank-one operators
\begin{eqnarray}
&& \cO_D^L \eqsp \tr(\hat Z^L) \, \nonumber \\
&& \cO_{\tilde E}^L \eqsp \tr(\hat Z^{L-2} \hat {\bar \Psi}_2 \hat {\bar \Psi}_2) \, , \label{oDEF} \\
&& \cO_F^L \eqsp \tr(\hat Z^{L-2} X X) \, , \qquad \ \,  \cO_{\tilde F}^L \eqsp \tr(\hat Z^{L-2} \bar X \bar X) \nonumber
\end{eqnarray}
to form the three classes of correlation functions
\beq
\cC_{A, \tilde A, D}^{L_1,L_2,L_3} \, \qquad \cC_{B, \tilde B, D}^{L_1,L_2,L_3} \, \qquad  \cC_{C, \tilde E, \tilde F}^{6,L_2,L_3}
\eeq
that we will study in this work. Moreover, when studying the effect of marginal deformations on structure constants in \ref{chapter4}, we will also look at the rank-one structure constants
\begin{align}
    \mathcal{C}_{F,\tilde F, D}^{L_1,L_2,L_3}~,
\end{align}
for both transversal and longitudinal excitations in $\cO_{F}$ and $\cO_{\tilde F}$, both at tree- and one-loop level.

\subsection{The coordinate Bethe ansatz for $psu(1,1|2)$ and nesting}

Barring for \emph{finite-size corrections} \cite{tba1,tba2,tba3,tba4,tba5,tba6}, which have to be computed by different means, the spin-chain description \cite{Minaza,beiStau1,beiStau2} of the planar higher-loop spectrum problem of the \nFour model is based on incorporating higher orders in the 't Hooft coupling $g^2 \eqsp g^2_\mathrm{YM}N/(16 \pi^2)$ into the Zhukowsky variables $x^\pm = u \pm i/2 + O(g^2)$. Here $N$ is the rank of the gauge group $SU(N)$ and the planar limit corresponds to $N\rightarrow\infty$, where $g^2$ is hold finite and small. In this article we are mainly interested in tree-level structure constants, and thus we solve the one-loop mixing problem to obtain the one-loop anomalous dimensions and leading-order eigenoperators.
Yet, there should be no obstruction to re-instating the Zhukowsky variables in order to recover the \emph{asymptotic part} of the coupling-constant dependence. Further, in the undeformed theory, finite-size corrections to structure constants can be computed by the \emph{gluing prescription} of the hexagon approach as advocated in \cite{BKV}. On the contrary, this last point is not obvious when marginal deformations are switched on.

The operator-mixing problem of planar $\cN=4$ SYM theory can be mapped to the spectral problem of an integrable spin-chain model \cite{Minaza,Beisert:2003yb,beiStau1,beiStau2}. Let us select the field $Z$ as the vacuum of the cyclic spin chains. The bosonic excitations $X, \bar X, Y, \bar Y,  D^{\alpha \dot \alpha}$ and the four fermions $\Psi^{3 \alpha}, \Psi^{4 \alpha}, \bar \Psi_1^{\dot \alpha}, \bar \Psi_2^{\dot \alpha}$ (so all in all eight bosonic and eight fermionic excitations) can occupy any lattice site. In the tensor product notation already employed in \eqref{secX}, \eqref{secBar} the full set \emph{magnons} is $\chi^A \otimes \chi^{\dot{A}}$, $\chi^A \eqsp \{ \phi^a, \psi^\alpha \}, \, \chi^{\dot{A}} \eqsp \{ \phi^{\dot{a}},  \psi^{\dot \alpha} \}$, where $a, \alpha\in \{1,2\}$ and $\dot{a}, \dot \alpha\in\{\dot 1,\dot 2\}$.

The magnons move along the chain with rapidities $u \eqsp 1/2 \cot(p/2)$, where $p$ is interpreted as a momentum. When a magnon overtakes another a rapidity-dependent phase factor arises, the $S$-\emph{matrix}. The total matrix is a tensor product of a \emph{left} and a \emph{right} factor, both given by the $ psu(2|2)$ invariant $(2|2) \otimes (2|2)$ $S$-matrix of \cite{beisertSu22} which we denote as $\mathbb{S}$. The entire action is normalised by
\beq
A_{jk}\equiv A(u_j,u_k) \eqsp \frac{u_j - u_k - i}{u_j - u_k +i} \, , \label{aMat}
\eeq
which corresponds to the $S$-matrix element for flipping two equal scalar excitations $\phi^a(u_j)$ and $ \phi^a(u_k)$ (no summation over $a$) on the left chain, and similarly $\phi^{\dot{a}}(u_j)$ and $\phi^{\dot{a}}(u_k)$ on the right chain. With the correct rapidities, this Bethe ansatz yields conformal eigenstates of the theory, i.e.\ states of well-defined anomalous dimension. 

\subsubsection{Coordinate Bethe ansatz for rank-one models}
The rapidities are determined by the Bethe equations and cyclicity constraint, and their explicit form depends on the type of excitations on the chain. Restricting to only one type of excitations, we find various \emph{rank-one sectors}: For scalars, fermions or derivatives  they are called an $su(2), \, su(1|1), \, sl(2)$ sector according to what lowering operator acts on the vacuum $Z$. The $su(2)$ case, say, with scalars $X$, is an instance of the famous Heisenberg spin chain which Bethe originally meant to diagonalise. Moving a magnon a spin-chain site ahead is accomplished by the shift
\beq
e^{i \, p} \eqsp \frac{u + \frac{i}{2}}{u - \frac{i}{2}} \, .
\eeq
Let there be $n$ scalars $X$ moving on a chain of total length $L$ (leaving $L-n$ instances of $Z$). Possible colour traces then take the form $\tr(X \, Z^i \, X \, Z^j \, X \ldots)$. Due to the cyclic nature of the trace this corresponds to a circular spin chain. Taking any one magnon once around chain we reproduce the original configuration, so that we require\footnote{Here we have $\phi^2$ on the left chain and $ \phi^{\dot{2}}$ on the right chain, both yielding factors $A_{jk}$ for each scattered magnon. The normalisation cancels one of these factors.}
\beq
e^{ i \, p_j \, L} \prod_{k \neq j}^n A_{jk}\eqsp 1 \, , \qquad j,k \, \in \{1 \ldots n\} \, . \label{betYa}
\eeq
Solving the coupled system of such \emph{Bethe--Yang equations} yields a discrete set of solutions for the rapidities. In addition, translation invariance along the cyclic chain implies $\sum p_i \eqsp 0$, or 
\beq
\prod_{j=1}^n \frac{u_j + \frac{i}{2}}{u_j - \frac{i}{2}} \eqsp 1 \, .
\label{zeroMom}
\eeq
Finally, the energy $E$ of the solution --- or in field-theory parlance, the planar one-loop anomalous dimension --- is given by
\beq
E \eqsp \sum_{i=1}^n \frac{1}{u_i^2 + \frac{1}{4}} \, .
\eeq
The last formula carries over to the nested Bethe ans\"atze discussed below; even though auxiliary rapidities are introduced, only the momentum carrying level-1 rapidities will contribute to the energy.

In order to construct an eigenstate of the one-loop mixing problem, first consider an open chain of length $L$. Suppose that $n$ scalar excitations occupy the lattice sites $m_j, \, j \, \in \, \{1 \ldots n\}$, while the vacuum $Z$ fills all other positions. The Bethe wave function is
\beq
\psi(L,n)\, =\!\! \sum_{m_1, \ldots, m_n \, = \, 1}^L \!\!\!\!\!\!\!\!\! e^{i\sum_{j=1}^n\, p_j \, m_j} \ \prod_{j < k} T_{jk} \ \, |m_1 \ldots m_j\dots m_k\dots m_n\rangle \, , \,\, T_{jk} \, = \, \Biggl\{ \begin{array}{ccl} m_j \, < \, m_k & : & 1 \\ m_j \, = \, m_k & : & 0 \\ m_j \, > \, m_k & : & A_{jk} 
\end{array} \label{wavSU2}
\eeq
This formula is valid \emph{off shell}, so even if the Bethe--Yang equations have not been imposed.
Substituting solutions of \eqref{betYa} and \eqref{zeroMom} into this ansatz, we find conformal eigenoperators. Unit norm is obtained by normalising as \cite{Korepin1,Korepin2}
\beq
\frac{\psi(L,n)}{\sqrt{L \, {\cal G} \, \prod_j (u_j^2 + \frac{1}{4}) \, \prod_{j \, < \, k} A_{jk}}} \, . \label{normWave}
\eeq
Note that the state does not depend on the ordering of the rapidities as long as the latter is equal in the wave function and the phase factor $\prod A_{jk}$. In the last formula, the \emph{Gaudin norm} is defined as
\beq
{\cal G} \, = \, \mathrm{Det} \ \phi_{jk} \, , \qquad \phi_{jk} \, = \, \frac{\partial  \log \left( e^{i \, p_j \, L } \prod_{l \, \neq \, j} A_{jl} \right)}{\partial \, u_k}~. \label{eq:Gaudin}
\eeq

\subsubsection{Multi-component wave functions for higher-rank models} \label{sec:MultiComp}

In order to generalise this construction to higher-rank cases, one can first look at the $tJ$ \emph{model}. This model is well-studied in the solid-state literature, and contains one type of bosonic and one type of fermionic excitations. We rely on the comprehensive discussion in the AdS/CFT context presented in \cite{beiStau1,beiStau2}. For excitations $X$ and $\Psi^4$, both sorts of magnons have $ \phi^{\dot{2}}$ on the right chain, thus scattering with each other picking up a phase $A_{jk}$ \eqref{aMat} which is offset by the common normalisation. Effectively, there is non-trivial scattering only on the left chain.

Suppose that we put the magnons $X(u_1), \Psi^4(u_2)$ onto a chain as in \eqref{wavSU2}. In principle, the structure of the wave function will be the same, but instead of the scattering
\beq
| \phi^2(u_1) \phi^2(u_2) \ra \rar A_{12} \, | \phi^2(u_2) \phi^2(u_1) \ra
\eeq
we now have to use
\beq
| \phi^2(u_1) \psi^2(u_2) \ra \rar G_{12} \, | \psi^2(u_2) \phi^2(u_1) \ra + H_{12} \, | \phi^2(u_2) \psi^2(u_1) \ra \, ,
\eeq
where $G_{12}$ and $H_{12}$ are elements from the S-matrix \cite{beisertSu22}. So there are two terms which are customarily called \emph{transmission} and \emph{reflection} because the flavours of the two excitations are transmitted through each other or they repel. For two magnons, formula \eqref{wavSU2} thus becomes
\begin{eqnarray}
\psi(L,\{X_1,\Psi_2\}) & = & \sum_{m_1 < m_2} e^{i (p_1 m_1 + p_2 m_2)} | X_1^{m_1} \Psi_2^{m_2} \ra +\nonumber  \\
&& \sum_{m_1 > m_2} e^{i (p_1 m_1 + p_2 m_2)}\left( G_{12} \, | \Psi_2^{m_2} X_1^{m_1} \ra + 
H_{12} \, | X_2^{m_2} \Psi_1^{m_1} \ra\right) \, .\label{wavXP}
\end{eqnarray}
To unclutter the notation we have omitted the flavour index of the fermion, instead we put the position of the excitation as a superscript. With the wave function \eqref{wavXP} alone, we cannot achieve periodicity because of the last term in which the flavours are exchanged. The solution to this problem is to introduce a second wave function 
\begin{eqnarray}
\psi(L,\{\Psi_1,X_2\}) & = & \sum_{m_1 < m_2} e^{i (p_1 m_1 + p_2 m_2)} | \Psi_1^{m_1} X_2^{m_2} \ra +  \nonumber \\
&& \sum_{m_1 > m_2} e^{i (p_1 m_1 + p_2 m_2)} \left( L_{12} \, | X_2^{m_2} \Psi_1^{m_1} \ra + K_{12} \, | \Psi_2^{m_2} X_1^{m_1} \ra \right) \, ,\label{wavPX}
\end{eqnarray}
where $L_{12}$ is the transmission and $K_{12}$ the reflection factor for the scattering of $\Psi^4$ with $X$, and to consider the sum
\beq
\psi^{1,1}(L) \eqsp g_{X \Psi} \ \psi(L,\{X_1,\Psi_2\}) + g_{\Psi X} \ \psi(L,\{\Psi_1,X_2\}) \, \label{psi11}
\eeq
with some as yet to be determined coefficients $g_{X \Psi}, \, g_{\Psi X}$. By imposing periodicity on the entire wave function, one obtains a $2 \times 2$ matrix of Bethe equations which fix the rapidities as well as the ratio of the coefficients.

The $G, H, K, L$ elements of the higher-loop scattering matrix $\mathbb{S}$ \cite{beiStau1,beiStau2,beisertSu22} reduce to the customary transmission and reflection amplitudes of the tJ model at lowest order in the coupling
\beq
G_{12} \eqsp L_{12} \eqsp \frac{u_1 - u_2}{u_1 - u_2 + i} \, , \qquad H_{12} \eqsp K_{12} \eqsp \frac{-i}{u_1 - u_2 + i}~.
\eeq
The example illustrates well how to generalise to higher number of magnons. We renounce on expanding on general formulae because wave functions with more excitations fan out considerably.

\subsubsection{Nested Bethe ansatz for higher-rank models}\label{sec:nestedpsu}

Let us now discuss the Bethe ansatz in the $psu(1,1|2)$ sector. First, we introduce the oscillator picture \cite{niklasDila,beiStau1,beiStau2,christophDila} where one can realise the elementary fields by fermionic creation operators $c^\dagger_I$, with flavour indices $I \, \in \, \{1 \ldots 4\}$ and obeying $\{c^\dagger_I,c_J\} \, = \, \delta_{IJ}$, and bosonic creation operators $a^\dagger_\alpha, b^\dagger_{\dot \alpha}$, with two-component spinor indices. The vacuum $Z$ and the fields $X,\Psi^4,\bar\Psi_1, D^{2\dot 2}$ of the first $psu(1,1|2)$ sector \eqref{secX} can be represented as
\begin{equation}
\begin{aligned}
&Z \eqsp c^\dagger_3 c^\dagger_4 | 0 \ra \, , &&  X \eqsp c^\dagger_2 c^\dagger_4 | 0 \ra\, ,&&\Psi^4 \eqsp \, a^\dagger_2 c^\dagger_4 | 0 \ra \, , && \bar \Psi_1 \, = \, b^\dagger_{\dot{2}} c^\dagger_2 c^\dagger_3 c^\dagger_4 | 0 \ra \, ,&& D^{2\dot 2} \, = \, a^\dagger_2 b_{\dot{2}}^\dagger c^\dagger_3 c^\dagger_4 | 0 \ra ~.
\end{aligned}
\end{equation}
The symmetry generators can also be expressed in the oscillator picture as
\begin{eqnarray}
\mathfrak{R}_3^2 \, \eqsp \phantom{-} c^\dagger_2 c_3  & : & Z \rar X \, , \nonumber \\
Q_2 \eqsp \phantom{-} a^\dagger_2 c_2 & : & X \rar \Psi^4 \, , \\
\bar Q^3 \, \eqsp - b^\dagger_{\dot{2}} c^\dagger_3 & : & X \rar \bar \Psi_1 \nonumber \, .
\end{eqnarray}
We have three lowering operators: $\mathfrak{R}_3^2$ creates the \emph{level-1 excitation} $X$ from the vacuum, while $Q_2$ and $\bar Q^3$ map $X$ to fermions. Note that the latter can only be created from $X$, and not directly from $Z$, so one speaks of \emph{level-2,3} or --- as the situation is symmetric --- rather \emph{left} and \emph{right} excitations, respectively. It is easy to verify that the fourth type of excitation $D^{2\dot 2}$ in this sector is created by the repeated action of these lowering operators, in particular
\begin{eqnarray}
\bar Q^3 \, Q_2 \, \mathfrak{R}_3^2 \, | Z \ra & = &  -D^{2 \dot 2} | Z \ra \, , \label{eq:leftrightDerivative}\\
\mathfrak{R}_3^2 \, \bar Q^3 \, Q_2 \, \mathfrak{R}_3^2 \, | Z \ra & = &  -D^{2 \dot 2} | X \ra \, . \nonumber
\end{eqnarray}
In rank-one sectors and the tJ model, the scalars and fermions behave like hard-core excitations, cf.\ $T_{jj}=0$ in \eqref{wavSU2}, as well as equations \eqref{wavXP} and \eqref{wavPX}. When there are both left and right excitations this is different: equations \eqref{eq:leftrightDerivative} show how a level-1, a left and a right excitation at the same site create a derivative of the vacuum $Z$, whereas a derivative of $X$ is obtained when two level-1 excitations, a left and a right magnon come together at the same site. Therefore, in the field-theory operator $\cO^l_B$ of \eqref{oABC}, there will only occur a single $X$ in the derivative terms. Our second $ psu(1,1|2)$ sector \eqref{secBar} works in the same way.\footnote{Note that also the Bethe ansatz for an $so(6)$ sector is strictly analogous in this respect: $\bar Z$ is created by a double level-1 excitation with a left and a right magnon on top. We thank M. Staudacher for a discussion about this point.}

One may endeavour to directly solve the matrix Bethe equations for this sector. However, more customarily the \emph{nested Bethe ansatz} is employed, which diagonalises the scattering matrix $\mathbb{S}$. The oscillator picture naturally points into this direction: let us again single out $\mathfrak{R}_3^2$ as the operator creating a level-1 excitation moving along a chain of vacua $Z$. This yields the previously discussed $su(2)$ sector with
\beq
S^{10} \eqsp e^{i \, p} \, , \qquad S^{11}_{jk}\equiv S^{11}(u_j,u_k) \eqsp A_{jk}
\eeq 
for moving the level-1 magnon over a vacuum, i.e.\ level-0, site and the scattering of two level-1 magnons. Note that in the following the superscripts indicate the level of the excitations involved in the scattering process, whereas the subscripts indicate the corresponding rapidities. Next, the left fermion can be created from $X$ by acting with $Q_2$. This excitation can move over a sub-chain composed of all the $X$. We attribute a secondary (auxiliary) rapidity $v$ to this motion. There is again a type of shift operator and a scattering matrix:
\beq
S^{21} \eqsp \frac{v - u  + \frac{i}{2}}{v - u - \frac{i}{2}} \, , \qquad S^{22}_{12}\equiv S^{22}(v_1,v_2) \eqsp -1 \, , \qquad f^{21}  \eqsp \frac{1}{v-u-\frac{i}{2}}~.
\eeq
The third quantity $f^{21}$ is the amplitude for creating a fermion with rapidity $v$ from an underlying $X$ with rapidity $u$. There is no such amplitude in the level-1 problem. Likewise, the right-excitation generator $\bar Q^3$ can transform $X$ into a right fermion. It moves along the sub-chain of $X$'s (also dubbed the \emph{level-1 vacuum}) with rapidity $w$. The corresponding elements of the Bethe ansatz are
\beq
S^{31} \eqsp \frac{w - u  + \frac{i}{2}}{w - u - \frac{i}{2}} \, , \qquad S^{33}_{12} \equiv S^{33}(w_1,w_2) \eqsp -1 \, , \qquad f^{31}  \eqsp \frac{1}{w-u-\frac{i}{2}}~.
\eeq
The left and right magnons are of fermionic nature and thus they scatter via $S^{23} \eqsp -1$. There is no $S^{20}$ and $S^{30}$ phase, such that the left and right magnons effectively move only on the level-1 vacuum consisting of the $X$'s. The derivation of the scattering matrices and creation amplitudes in the $S$-matrix formalism (without the nesting of excitations) is reviewed in many places in the literature and will not be repeated here. We would like to point the reader's attention to \cite{beiStau1,beiStau2,beisertSu22}.

A peculiarity of the ansatz (that we have not seen elsewhere in the literature) is the existence of a second type of creation amplitude for the quadruple excitation $D X$. A left and a right fermion at different sites of the chain are created with the product of amplitudes $f^{21}(v,u_1) f^{31}(w,u_2)$. The polynomial order of this product is $O(u^{-2})$. For the excitation $DX$ there is one left excitation and one right excitation located on top of a double level-1 excitation, and thus we would expect the product $f^{21}(v,u_1) f^{21}(v,u_2) f^{31}(w,u_1) f^{31}(w,u_2)$ which is of order $O(u^{-4})$. Hence for reasons of homogeneity there should be a quadratic numerator. The right guess\footnote{Barring for the overall sign this is similar to the $so(6)$ Bethe ansatz, for which we originally derived this creation amplitude by matching on the dilatation operator \cite{niklasDila,christophDila} eigenstates with $n_1 \eqsp 2, \, n_2 \eqsp n_3 \eqsp 1, \, L \eqsp 2 \ldots 9$ and $n_1 \eqsp 3, \, n_2 \eqsp n_3 \eqsp 1, \, L \eqsp  5,6$. Here the number of level-1 excitations is $n_1$, and the number of level-2,3 excitations is $n_{2,3}$. In the latter two cases there are have non-vanishing secondary rapidities.} seems to be 
\beq
f^4(\{u_1,u_2\},v,w) \eqsp - f^{21}(v,u_1) f^{21}(v,u_2) f^{31}(w,u_1) f^{31}(w,u_2) (u_1-u_2)(u_1-u_2-i) \, . \label{nested4}
\eeq
Because the formula is not $u_1 \, \leftrightarrow \, u_2$ symmetric, it is important to respect the exact ordering of the rapidities $u_1,u_2$ with which the level-1 magnons were originally put on the chain.
 
In the Bethe equations, the nested structure becomes apparent: the level-1 magnons are given rapidities $u_i$ with $i \, \in \, \{1 \ldots n_1\}$ and $n_1 \eqsp n_X + n_\Psi + n_{\bar \Psi}$ referring to the respective numbers of excitations.
On the level-1 vacuum we have $n_2=n_\Psi$ left excitations and $n_3=n_{\bar\Psi}$ right excitations. We can once again derive Bethe--Yang equations by demanding that the original configuration is identically reproduced upon taking one excitation around the chain:
\begin{eqnarray}
1 & = & \prod_{j =1}^{n_1} e^{i p_j} \, , \nonumber \\
1 & = & e^{i p_i L} \, \prod_{j \neq i, \, j = 1}^{n_1} S^{11}(u_i,u_j) \, \prod_{j = 1}^{n_2} \frac{1}{S^{21}(v_j,u_i)} \, \prod_{j=1}^{n_3} \frac{1}{S^{31}(w_j,u_i)} \, , \nonumber \\
1 & = & \prod_{j = 1}^{n_1} S^{21}(v_i,u_j)\, , \nonumber\\
1 & = & \prod_{j = 1}^{n_1} S^{31}(w_i,u_j)  \, .\label{eq:BetheSU112}
\end{eqnarray}
We have omitted $S^{22}, \, S^{23} \, S^{33}$ which would have to be compensated by extra minus signs to obtain the correct solution to the fermionic Bethe equations.

For a given ordering of the rapidities, the nested Bethe ansatz yields a definite wave function. Even off-shell (i.e.\ prior to imposing the Bethe--Yang equations) we can equate this with the multi-component wave function although the latter contains far less coefficients than there are distinct ket states.

\subsection{Hybrid formalism for higher-rank operators} \label{sec:Hybrid}

Matching the wave function of the nested Bethe ansatz (discussed in Sec.\ \ref{sec:nestedpsu}) on the multi-component wave function (discussed in Sec.\ \ref{sec:MultiComp}), one can extract the coefficients for every component, e.g.\ $g_{X\Psi}$ and $g_{\Psi X}$ of \eqref{psi11} for the tJ model. While the entire wave function can be very complicated, these coefficients will turn out to have a simple and systematic structure. By a literature search or else in future work it would be nice to arrive at general expressions.

In the following we discuss the extraction of the multi-component wave-function coefficients from the nested Bethe ansatz for the operators in \eqref{oABC}. Note that all primary operators studied in this work assume unit norm if scaled as in \eqref{normWave}. Albeit ${\cal G}$ from \eqref{eq:Gaudin} is now generalised to the \emph{full Gaudin determinant}, where $-i \log$ of every Bethe equation is differentiated w.r.t.\ to all rapidities $\{u_i,v_j,w_k\}$. 

\subsubsection{Matching the operator $\cO_A^L$}

We begin with the operator $\cO_A^L$ which contains two excitations $X$ and $\Psi^4$ on a vacuum of $L-2$ $Z$'s, and similarly for $\cO_{\tilde A}^L$ with excitations $\bar{X}$ and $\bar\Psi_1$. The partial wave functions of the multi-component ansatz are constructed with initial momentum ordering $\{u_1,u_2\}$,  and likewise for the level-1 roots in the nested approach. Matching the coefficients of both wave functions, we find only two independent conditions with the solution
\beq
g_{\Psi X} \, = \, f^1(v) , \qquad g_{X \Psi} \, = \, f^2(v) \, , \label{superClean}
\eeq
where
\beq
f^k(v) \, = \, f^{21}(v,u_k) \prod_{j=1}^{k-1} S^{21}(v,u_j) \, . \label{whatF}
\eeq
The pattern is very transparent and lends itself to generalisation! Note that the multi-component wave function with coefficients \eqref{superClean} now depends not only on the two rapidities $u_1$ and $u_2$ of the bosonic and fermionic excitations, but also the auxiliary root of the nested Bethe ansatz. 

\subsubsection{Matching the operator $\cO_C^L$}

The operator $\cO_C^L$ is constructed from two bosonic excitations $X$ and two fermionic excitations $\Psi^4$ on a vacuum of $L-4$ $Z$'s. These four excitations can appear in six different initial sequences. For $L=6$ and disregarding cyclic invariance, there are 90 distinct ket states. Nevertheless, substituting some random numerical values for the rapidities we can quickly check that the system of equations has a unique solution for the six unknown coefficients. On the other hand, a full analytic solution for general $L$ is quite hard to obtain because there are four level-1 rapidities so that we cannot expect to see the equivalence of the various equations without appealing to the Yang--Baxter equation. The situation is further complicated by the presence of the two auxiliary rapidities. However, the length $L=6$ primary state with anomalous dimension 10 is not part of a degenerate pair and therefore has symmetric root distribution (the exact values are given in \eqref{rootsL6G10}). We can thus reduce the complexity of the problem by identifying $u_4 \eqsp -u_1, \, u_3 \eqsp -u_2, \, v_2 \eqsp -v_1$ which is sufficient to guide {\tt Mathematica} to a solution. Let $j,k \, \in \, \{1 \ldots 4\}$ denote the position of the level-2 excitations on the level-1 vacuum $X(u_1) X(u_2) X(u_3) X(u_4)$. We find 
\beq
g_{jk} \eqsp f^j(v_1) f^k(v_2) - f^j(v_2) f^k(v_1)~. \label{superCleanC}
\eeq
Furthermore, numerically this equation is confirmed for general off-shell rapidities, too. Combined with the later result \eqref{ansFB1}, these results for the $g$ coefficients suggest a tensor-product structure with antisymmetrisation on the auxiliary rapidities. Note that these results are very close to the material of Appendix F in \cite{nestedHexagon} where it is observed that some related wave functions are \emph{Slater determinants}. Combined with the later result \eqref{ansFB1}, these results for the $g$ coefficients suggest a tensor-product structure with antisymmetrisation on the auxiliary rapidities.

\subsubsection{Matching the operator $\cO_B^L$} \label{secOB}

To fix the multi-component wave-function for operator $\cO_B^L$ (and similarly  $\cO_{\tilde B}^L$), we need to determine the four coefficients $g_{\Psi \bar \Psi}$,  $g_{\bar \Psi \Psi}, \, g_{D X}, \, g_{X D}$. The length $L=3$ problem has 15 ket states. Equating with the nested Bethe ansatz, we can easily solve the linear system arising from the twelve cases without $D X$:
\beq
\begin{aligned}
&g_{\Psi \bar \Psi} \eqsp f^1(v) f^2(w) \, , \qquad &&g_{\bar \Psi \Psi} \eqsp - f^1(w) f^2(v) \, , \\
&g_{D X} \eqsp f^1(v) f^1(w) \, , \qquad &&g_{X D} \eqsp f^2(v) f^2(w) \, . \label{ansFB1}
\end{aligned}
\eeq
Here we have used the symbol $f^k$ also for the right magnon because $f^{31} \eqsp f^{21}, \, S^{31} \eqsp S^{21}$ as functions of two general arguments. At first sight, the solution fails for the remaining three kets $| (DX) Z Z \ra, |Z (DX) Z \ra, | Z Z (DX) \ra$. The missing piece of the ansatz is in fact a creation amplitude for the quadruple excitation
\beq
f^0(u_1,u_2) \, = \, c \ \frac{u_1 - u_2}{u_1 - u_2 - i} \label{straight4}
\eeq
caused by \eqref{nested4} from the nested ansatz. Again, it is vital to keep track of the order of the rapidities because the amplitude is not exchange-symmetric. In the last formula, $c$ is a sign depending on which wave function the quadruple excitation derives from:
\beq
1 \eqsp - \, c_{\Psi \bar \Psi} \eqsp c_{\bar \Psi \Psi} \eqsp c_{D X} \eqsp c_{X D}~. \label{ansFB2}
\eeq
The ansatz \eqref{ansFB1}, \eqref{straight4}, \eqref{ansFB2}  solves the length 4, 5 problems, too.

\subsection{Entangled states and partitions for the hexagon} \label{secEntangled}

In the hexagon approach to structure constants the three-point function is cut into two halves. For the field theory/integrability description this implies that the single-trace operators at the outer points --- in other words the Bethe states  --- are split, too. In \cite{BKV}, and indeed in most other computations in the literature, this procedure is done for rank-one sector states. As a first example, consider an operator of length $L$ carrying two excitations from a single $su(2)$ sector with momenta $p_1$ and $p_2$. Cutting it into two substates of lengths $l_1$ and $l_2$, with $L=l_1+l_2$, the excitations can end up on either subchain. We have
\begin{align}
\psi(L, \{X_1,X_2\}) &= \psi(l_1,\{X_1,X_2\}) \, \psi(l_2, \{\}) + e^{i \, p_2 \, l_1} \,\psi(l_1,\{X_1\}) \, \psi(l_2, \{X_2\})\, + \label{entSU2} \\ & e^{i \, p_1 \, l_1} A_{12}  \ \psi(l_1,\{X_2\}) \, \psi(l_2, \{X_1\}) +  e^{i (p_1+p_2) l_1} \, \psi(l_1,\{\}) \, \psi(l_2, \{X_1,X_2\})~. \nonumber
\end{align}

Considering each of the multi-component wave function in the $psu(1,1|2)$ sector separately, the construction works in a similar way.
For example for $\cO_A^{L}$ we have
\begin{eqnarray}
\psi(L, \{X_1,\Psi_2\}) & \equiv & \psi(l_1,\{X_1,\Psi_2\}) \, \psi(l_2, \{\}) + e^{i \, p_2 \, l_1} \,\psi(l_1,\{X_1\}) \, \psi(l_2, \{\Psi_2\}) + \label{psi12XP} \nonumber \\ & & e^{i \, p_1 \, l_1} \bigl[ G_{12} \ \psi(l_1,\{\Psi_2\}) \, \psi(l_2, \{X_1\}) +  H_{12} \ \psi(l_1,\{X_2\}) \, \psi(l_2, \{\Psi_1\}) \bigr] + \nonumber \\ && e^{i (p_1+p_2) l_1} \, \psi(l_1,\{\}) \, \psi(l_2, \{X_1,\Psi_2\}) 
\end{eqnarray}
and
\begin{eqnarray}
\psi(L, \{\Psi_1,X_2\}) & \equiv & \psi(l_1,\{\Psi_1,X_2\}) \, \psi(l_2, \{\}) + e^{i \, p_2 \, l_1} \,\psi(l_1,\{\Psi_1\}) \, \psi(l_2, \{X_2\}) + \label{psi12PX} \nonumber \\ & & e^{i \, p_1 \, l_1} \bigl[ L_{12} \ \psi(l_1,\{X_2\}) \, \psi(l_2, \{\Psi_1\}) +  K_{12} \ \psi(l_1,\{\Psi_2\}) \, \psi(l_2, \{X_1\}) \bigr] + \nonumber \\ && e^{i (p_1+p_2) l_1} \, \psi(l_1,\{\}) \, \psi(l_2, \{\Psi_1,X_2\}) \, .
\end{eqnarray}
In the total wave function $g_{X\Psi}\psi(L, \{X_1,\Psi_2\}) +g_{\Psi X}\psi(L, \{\Psi_1,X_2\})$ we see that two of the split wave functions occur twice, in particular
\begin{eqnarray}
\psi^{1,1} & = & \ldots + \psi(l_1,\{\Psi_2\}) \, \psi(l_2, \{X_1\}) e^{i \, p_1 \, l_1} \, \bigl[ g_{X \Psi} \, G_{12} + g_{\Psi X} \, K_{12} \bigr] + \label{eightpartys}\\
&& \phantom{\ldots +} \, \psi(l_1,\{X_2\}) \, \psi(l_2, \{\Psi_1\}) \, e^{i \, p_1 \, l_1} \, \bigl[ g_{X \Psi} \, H_{12} + g_{\Psi X} \, L_{12} \bigr] + \ldots \nonumber 
\end{eqnarray}
The terms in the two square brackets factor in a suggestive way and applying the results from the matching with the nested Bethe ansatz of \eqref{superClean} we find the \emph{splitting factors} 
\begin{equation}
\begin{aligned}
&e^{i \, p_1 \, l_1} \left[ g_{X \Psi} \, G_{12} + g_{\Psi X} \, K_{12} \right] \eqsp  \, e^{i \, p_1 \, l_1} f^{21}(v_1,u_1)  \, S^{21}(v_1,u_2) \, S^{11}(u_2,u_1) , \\ & e^{i \, p_1 \, l_1} \left[ g_{X \Psi} \, H_{12} + g_{\Psi X} \, L_{12} \right] \eqsp e^{i \, p_1 \, l_1} f^{21}(v_1,u_2) \, S^{11}(u_2,u_1)~, \label{suggestiveF}
\end{aligned}
\end{equation}
which inherit a dependence on the auxiliary Bethe roots from the coefficients $g_{X\Psi}$ and $g_{\Psi X}$.
In the literature the terms of the multi-component wave function are often re-grouped by the order in which the flavours of the excitations appear on the chain, cf.\ \cite{beiStau1,beiStau2}. In our example we should thus cross-attribute the transmission terms, so the one with $L_{12}$ in \eqref{psi12PX} and the one with $G_{12}$ in \eqref{psi12XP} as belonging to new wave functions with magnon ordering $\{X, \Psi\}$ and $\{\Psi, X\}$, respectively. We do not adopt this point of view in this work because the notion of entangled states would be blurred. Yet, having constructed the various entangled states, it is perfectly natural to sort by inequivalent partitions and to factor their total splitting amplitudes in order to minimise the number of terms. In this way, thanks to \eqref{suggestiveF}, we find eight partitions for $\cO_A^{l_1+l_2}$ as opposed to four for the two-excitation $su(2)$ case. This formalism puts the added complexity of higher-rank sectors simply into forming larger sets of partitions.

\section{Structure constants in \nFour SYM} \label{chapter3}

We now move on to the computation of three-point functions of the operators \eqref{oABC} and \eqref{oDEF} and the associated structure constants. As long as only one higher-rank operator is put into a three-point function, our approach from Sec.\ \ref{sec:Hybrid} is not very different from what is done in \cite{nestedHexagon}. Yet, now the information provided by the auxiliary Bethe roots resides in the splitting factors, while the hexagon computations themselves will be done with the original formalism as defined in \cite{BKV}. As a consequence, we need not worry about implementing the crossing prescriptions for hexagons within the nested Bethe ansatz. This empowers us to consider correlation functions with more than one higher-rank operator.

We will not provide more detail on the derivation of the amplitudes since the method is very well explained in the original work \cite{BKV}.

\subsection{Correlators of the form $\cC_{A, \tilde A, D}$}

Solving the Bethe equations \eqref{eq:BetheSU112} for one bosonic and one fermionic excitation, one finds 
$u_2 \, = \, - u_1, \, v \, = \, 0$ (or $w \, = \, 0$, respectively) at any length $L$. Combined with \eqref{superClean} and \eqref{whatF}, this yields the very simple coefficients
\beq
g_{X \Psi} \, = \, \frac{1}{u_1+\frac{i}{2}} \, = \, - g_{\Psi X} \label{amusing}
\eeq
which are immediately rendered real when dividing out the standard phase $\sqrt{S^{11}(u_1,-u_1)}$ from \eqref{normWave}.

The shortest operators in the series are characterised by their rapidities $u_1$ and the one-loop anomalous dimension $E$:
\vskip 0.2 cm
\begin{center}
\begin{tabular}{c|c|c} 
$L$ & $u_1$ & $E$  \\
\hline
3 & $\frac{1}{\sqrt{12}}$ & 6 \\[2 mm]
4 & $\frac{1}{2}$ & 4 \\[2 mm]
5 & $\frac{1}{2} \sqrt{1 \pm \frac{2}{\sqrt{5}}}$ & $5 \mp \sqrt{5}$
\end{tabular}
\end{center}
This is valid independently of the choice of magnon pair. Since all the excitations are transversal, we must choose conjugate operators at points 1 ($a=0$) and 2 ($a=1$) in the three-point function. For definiteness, let us pick $\{X, \Psi^4\}$ and $\{\bar X, \bar \Psi_2\}$, respectively. 

Cutting the three-point function into two hexagonal patches, we can use the hexagon form factor to evaluate each one of these patches. To obtain the structure constant one has to sum over all partitions of the excitations on the two hexagons.
We will denote the hexagon form factor by $\langle \mathfrak{h} | \alpha_1, \alpha_2, \alpha_3 \rangle$, where the sets $\alpha_1,\alpha_2,\alpha_3$ correspond to the excitations on the three physical edges of one hexagon. The excitations can be brought to the same edge using crossing transformations $\gamma$. We denote the edges (i.e.\ the sets $\alpha_1,\alpha_2,\alpha_3$) in the hexagon amplitudes as the $4\gamma$, $2\gamma$, and $0\gamma$ edges, respectively.

According to equations \eqref{psi12XP} - \eqref{eightpartys} each entangled state has eight partitions. The splitting amplitudes are defined by \eqref{suggestiveF} and \eqref{amusing}. 
Due to the transversal nature of the magnons, only the following 16 out of 64 (products of) hexagon amplitudes are non-vanishing (the two products in the first line of the next formula can each occur in four versions distinguished by the distribution of the momenta):
\begin{eqnarray}
&& \langle \mathfrak{h} | \{\Psi\}, \{\bar \Psi\}, \{\} \rangle \langle \mathfrak{h} |\{X\}, \{\}, \{\bar X\} \rangle,  \quad \langle \mathfrak{h} |\{\Psi\}, \{\}, \{\bar \Psi\} \rangle \langle \mathfrak{h} |\{X\}, \{\bar X\}, \{\} \rangle, \nonumber \\
&& \langle \mathfrak{h} |\{\Psi, X\}, \{\}, \{\bar \Psi, \bar X\} \rangle, \quad \langle \mathfrak{h} |\{\Psi, X\}, \{\}, \{\bar X, \bar \Psi\} \rangle, \label{hexAmpsAAD} \\
&& \langle \mathfrak{h} |\{\Psi, X\}, \{\bar \Psi, \bar X\}, \{\} \rangle, \quad \langle \mathfrak{h} |\{\Psi, X\}, \{\bar X, \bar \Psi\}, \{\} \rangle, \nonumber \\
&& \langle \mathfrak{h} |\{X, \Psi\}, \{\}, \{\bar \Psi, \bar X\} \rangle, \quad \langle \mathfrak{h} |\{X, \Psi\}, \{\}, \{\bar X, \bar \Psi\} \rangle, \nonumber \\
&& \langle \mathfrak{h} |\{X, \Psi\}, \{\bar \Psi, \bar X\},\{\} \rangle, \quad \langle \mathfrak{h} |\{X, \Psi\}, \{\bar X, \bar \Psi\}, \{\} \rangle \, . \nonumber
\end{eqnarray}
For simplicity, we omitted the empty hexagon $\langle \mathfrak{h} | \{\}, \{\}, \{\} \rangle \, = \, 1$ multiplying the last eight cases. Likewise, the flavour indices on $\Psi^4, \, \bar \Psi_2$ are dropped.\footnote{A second choice would be $\{X, \bar \Psi_1\}$ at point $a=0$ and $\{\bar X, \Psi^3\}$ at point $a=1$. The hexagon amplitudes with fermions then simply switch sign so that we find an overall minus sign. This is reproduced by the field-theory calculation due to a flipped direction of the fermion propagator.}.
We refer the reader to the original literature \cite{BKV} for more details.

The hexagon amplitudes have real poles for coincident rapidities at points 1 and 2, the so-called \emph{particle-creation poles}. This is an essential feature of the formalism that appears to be lost in the nested-hexagon approach of \cite{nestedHexagon} (at least in the version of Appendix F of this work), but that we aspire to preserve in this formalism. On the other hand, these poles need to be regularised or factored out when dealing with degenerate rapidity sets. In previous work \cite{colourDressed,doubleTorus} it became apparent that the poles most easily factor out when the partitioning of the two Bethe states is not started on the same hexagon, so that the magnons can all be brought over the connecting edge between the operators at points 1 and 2. Its width is determined by the lengths $L_{1,2,3}$ of the three operators, so $l_{12} \, = \, (L_1 + L_2 - L_3)/2$. Fortunately, the problem at hand is simple enough to allow for analytic factorisation at width $l_{12} \eqsp 0 \ldots 5$ of the connecting edge. The cases $l_{12}=0$ and $1$ vanish, the others have a universal prefactor
\beq
p(l_{12}) \eqsp \frac{(u_1-u_2)(u_3-u_4) \, f^{21}(v_1,u_1) f^{21}(v_1,u_2) \, f^{31}(w_3,u_3)f^{31}(w_3,u_4)}{ (u_1-u_2+i)(u_3-u_4+i) \prod_{i=1}^4 (u_i-\frac{i}{2})^{l_{12}}} \label{pref}
\eeq
(we have employed rapidities $u_1,u_2,v_1$ at point 1 and $u_3,u_4,w_3$ at point 2) times a fairly complicated polynomial $q(l_{12})$ of maximal order $4 \, l_{12} + 2$ in all rapidities, but maximally $l_{12}$ in each of the level-1 rapidities $u$ and linear in the auxiliary ones $v$ and $w$. That the particle-creation poles disappear for generic rapidities is a sharp test for the validity of our procedure. 
Imposing the properties of the solution $u_2 \eqsp -u_1, \, v_1 \eqsp 0$ etc. the first three polynomials shrink to
\begin{eqnarray}
q(2) & = &  - 8 \, K_{13} \, , \qquad \qquad \qquad \qquad \qquad \qquad K_{13} \eqsp (u_1^2+1/4)(u_3^2+1/4) \, , \nonumber \\
q(3) & = & - (3 - 4 \, u_1^2 - 4 \, u_3^2 + 48 \, u_1^2 \, u_3^2) \, K_{13} \, , \\
q(4) & = &  -(1/8) (5 - 4 \, u_1^2 - 4 \, u_3^2 + 16 \, u_1^2 \, u_3^2) (1 - 4 \, u_1^2 - 
   4 \, u_3^2 + 80 \, u_1^2 \, u_3^2) \, K_{13} \, , \nonumber
\end{eqnarray}
(we do not display $q(5)$ as it is more involved), whereas 
\beq
\frac{p(l_{12})}{\sqrt{S^{11}(u_1,u_2) S^{11}(u_3,u_4)} }\rar \frac{u_1 u_3}{K_{13}^{(l_{12} + 3/2)}} \, .
\eeq

Let us come back to the evaluation of $\mathcal{C}_{A,\tilde{A},D}^{L_1,L_2,L_3}$. For a given width of the connecting edge $l_{12}$, the vacuum at the third point must have length $L_3 \eqsp L_1 + L_2 - 2 \, l_{12}$. Dropping a global minus sign, for $L_3 \eqsp 0$ the hexagon results normalised by the full Gaudin determinant ${\cal G}$ are orthonormal. 
On the other hand, if $L_3 > 1$ the hexagon result, made real by the standard phases and normalised by the root of the full Gaudin determinants of the two non-trivial states, should reproduce $1/(L_1 L_2 L_3)$ times the leading-$N$ part of the corresponding connected free field-theory correlators. This is indeed the case!
Ordering the four operators in the table above as $\cO_A^{3}, \cO_A^{4}, \cO_A^{5+},\cO_A^{5-}$ we can build $4 \times 4$ three-point functions by adding an appropriate vacuum operator, resulting in a matrix of structure constants. For $l_{12} \eqsp 2$ we find
\beq
- \left(
\begin{array}{cccc}
 \sqrt{2} & \sqrt{3} & \sqrt{\frac{2}{5} \left(5+\sqrt{5}\right)} & \sqrt{\frac{2}{5}
   \left(5-\sqrt{5}\right)} \\[1 mm]
 \sqrt{3} & 2 & \sqrt{\frac{1}{2} \left(5+\sqrt{5}\right)} & \sqrt{\frac{1}{2}
   \left(5-\sqrt{5}\right)} \\[1 mm]
 \sqrt{\frac{2}{5} \left(5+\sqrt{5}\right)} & \sqrt{\frac{1}{2} \left(5+\sqrt{5}\right)}
   & \sqrt{\frac{3}{5} \left(3+\sqrt{5}\right)} & \sqrt{\frac{6}{5}} \\[1 mm]
 \sqrt{\frac{2}{5} \left(5-\sqrt{5}\right)} & \sqrt{\frac{1}{2} \left(5-\sqrt{5}\right)}
   & \sqrt{\frac{6}{5}} & \sqrt{\frac{3}{5} \left(3-\sqrt{5}\right)} \\
\end{array}
\right) \label{matA2}
\eeq
whereas for $l_{12} \eqsp 3$ we find (an asterisk denotes an inexistent correlator)
\beq
- \left(
\begin{array}{cccc}
 * & * & \sqrt{5-\sqrt{5}} & \sqrt{5+\sqrt{5}} \\[1 mm]
 * & 2 \sqrt{2} & \sqrt{\frac{6}{5} \left(5+\sqrt{5}\right)} & \sqrt{\frac{6}{5}
   \left(5-\sqrt{5}\right)} \\[1 mm]
 \sqrt{5-\sqrt{5}} & \sqrt{\frac{6}{5} \left(5+\sqrt{5}\right)} & 3 +\frac{1}{\sqrt{5}} & \frac{2}{\sqrt{5}} \\[1 mm]
 \sqrt{5+\sqrt{5}} & \sqrt{\frac{6}{5} \left(5-\sqrt{5}\right)} & \frac{2}{\sqrt{5}} &
   3 -\frac{1}{\sqrt{5}} \\
\end{array}
\right)~.
 \label{matA3}
\eeq
The $l_{12} \eqsp 4$ connected correlators only exist for $L_1=L_2 \eqsp 5$ and we find the matrix
\beq
3 \sqrt{2} \left( \begin{array}{cc} 1 & 0 \\ 0 & 1 \end{array} \right) \, .
\eeq

\subsection{Correlators of the form $\cC_{B, \tilde B, D}$}

As the operators $\cO_B$ and $\cO_{\tilde B}$ each have a left and a right fermion, both have a $v$ and a $w$ rapidity. Again, we consider primary states of length 3 \ldots 5. The rapidities obey $u_2 \eqsp -u_1, \, v \eqsp 0 \eqsp w$. The states are distinguished by
\vskip 0.2 cm
\begin{center}
\begin{tabular}{c|c|c} 
$L$ & $u_1$ & $E$  \\
\hline
3& $\frac{1}{2}$ & 4 \\[2 mm]
4 & $\sqrt{1 \pm \frac{2}{\sqrt{5}}}$ & $5 \mp \sqrt{5}$ \\[2 mm]
5 & $\frac{\sqrt{3}}{2}$ & 2 \\[2 mm]
5 &$ \frac{1}{\sqrt{12}}$ & 6 
\end{tabular}
\end{center}

In parts, we see the same rapidities as those in the $\cO_A$ case from above. This is a consequence of \nFour supersymmetry: the commutator term in the fermion equation of motion as well as the variation of the covariant derivative augment the length by one unit:
\beq
Q_1 \, \Psi^4 \eqsp - 2 \, i \, g_\mathrm{YM}\,  [X,Z] \, , \qquad Q_1 \, D(\bullet) \eqsp Q_1 (\partial \bullet - i \, g_\mathrm{YM} \,  [A, \bullet]) \eqsp 2 \, i  \, g_\mathrm{YM}  \, [\bar \Psi_1, \bullet ]~. \label{dQ1}
\eeq 
The second equality holds for arguments that do not themselves transform under $Q_1$. Similarly, $\bar Q^4 \, \bar \Psi_1 \, \propto \, [X,Z]$ and the variation of the covariant derivatives brings out a commutator with $\Psi^4$. Note that the extra $g_\mathrm{YM}$ in \eqref{dQ1} implies that these are shortening directions in the free theory, which indicates multiplet splitting. 
Both supersymmetries $Q_1$ and $\bar Q^4$ are in the magnon group whence they do not act on the vacuum. One can move from one Bethe state to another via
\beq
\cO_B^L \ \stackrel{\bar Q^4}{\rightarrow} \ \cO_A^{L+1} \ \stackrel{Q_1}{\rightarrow}  \ \cO_F^{L+2}~, \label{susyChain1}
\eeq
up to some factors of proportionality, and likewise
\beq
\cO_B^L \ \stackrel{Q_1}{\rightarrow} \ \cO_{\check A}^{L+1} \ \stackrel{\bar Q^4}{\rightarrow} \ \cO_F^{L+2} \, . \label{susyChain2}
\eeq
This might seem surprising at first sight, as in the Bethe ansatz the sign of a descendant --- so something obtained by acting on a Bethe state with a symmetry generator --- is usually an infinite rapidity (or more than one). The reason that we do not see this here is that the $psu(1,1|2)$ group acting on $\{X,Z,\Psi^4,\bar \Psi_1\}$ contains the supersymmetries $Q_2,Q_3,\bar Q^2, \bar Q^3$ and not those in the last two formulae. Thus, within our Bethe ansatz the operators $\cO_B, \cO_A, \cO_F$ all are primaries. The situation in the second transversal $psu(1,1|2)$ sector is the same with $\{2,3\} \, \leftrightarrow \{1,4\}$. 

The hexagon computation in this section is more involved, but in spirit very similar to that of the last section. First, the Bethe states are more complicated because the scattering can trade a fermion/anti-fermion pair for a scalar and a derivative and vice versa. Yet, letting each of the four wave functions introduced in Section \ref{secOB} evolve freely, entangled states can be obtained just like before. There are 16 partitions at both points 1,2.  Similar to \eqref{suggestiveF} all splitting factors become simple products of momentum factors, creation amplitudes, and scattering matrices of the nested Bethe ansatz (the creation amplitude for the quadruple excitation does not appear).

Out of the 256 products of hexagon amplitudes only 48 are non-vanishing. Since all magnons are transversal, one would naively expect that only hexagon amplitudes containing conjugate pairs of excitations can be non-vanishing. But in fact, there are some non-vanishing hexagons with excitations $\{D, X, \Psi, \bar \Psi\}$, too. We will again start the partitioning, say, for point 1 on the back hexagon and for point 2 on the front hexagon, so bringing all magnons over the connecting edge of width $l_{12}$. Putting everything together we can factor out the particle-creation poles subject to introducing some fermion signs:
\begin{itemize}
\item Bringing one fermion of the second operator from $0 \gamma$ to $2 \gamma$ (i.e.\ from the front to the back hexagon), we have to choose the branch
\beq
\eta \xrightarrow[]{2\gamma} \frac{-i \, \eta}{\sqrt{x^-} \sqrt{x^+}} \, , \qquad \eta \eqsp \sqrt{i(x^- - x^+)}
\label{sigGamma}
\eeq
for crossing the fermion normalisation from the $S$-matrix of \cite{beisertSu22}. This also applies to the example in the last section. We emphasise that crossing can only be defined using the Zhukowsky variables as it does not commute with the coupling expansion. Only after the crossing can we expand in the 't Hooft coupling.
\item When both fermions of the second operator are brought to the $2 \gamma$ edge of the back hexagon, there is a minus sign. Similar effects have been seen in \cite{jaoThiago}.
\end{itemize}
As before, the disconnected parts ($L_3=0$) evaluate to the norm 1. For the connected part at edge width
$l_{12} \eqsp 1$, we find (this exists also in free field theory due to the $\tr(Z^{L-1} DX)$ terms)
\beq
\left(
\begin{array}{ccccc}
 1 & \frac{1}{2}\sqrt{3+\sqrt{5}} & \frac{1}{2} \sqrt{3-\sqrt{5}} & \sqrt{\frac{3}{2}} & \frac{1}{\sqrt{2}} \\[1 mm]
 \frac{1}{2}\sqrt{3+\sqrt{5}} & \frac{1}{5} \sqrt{3 \left(7+3 \sqrt{5}\right)} & \frac{\sqrt{6}}{5} &
   \frac{1}{2} \sqrt{\frac{7}{5} \left(3+\sqrt{5}\right)} & \frac{1}{2} \sqrt{\frac{7}{15}
   \left(3+\sqrt{5}\right)} \\[1 mm]
 \frac{1}{2}\sqrt{3-\sqrt{5}} & \frac{\sqrt{6}}{5} & \frac{1}{5} \sqrt{3 \left(7-3 \sqrt{5}\right)} &
   \frac{1}{2} \sqrt{\frac{7}{5} \left(3-\sqrt{5}\right)} & \frac{1}{2} \sqrt{\frac{7}{15}
   \left(3-\sqrt{5}\right)} \\[1 mm]
 \sqrt{\frac{3}{2}} & \frac{1}{2} \sqrt{\frac{7}{5} \left(3+\sqrt{5}\right)} & \frac{1}{2} \sqrt{\frac{7}{5}
   \left(3-\sqrt{5}\right)} & \sqrt{2} & \sqrt{\frac{2}{3}} \\[1 mm]
 \frac{1}{\sqrt{2}} & \frac{1}{2} \sqrt{\frac{7}{15} \left(3+\sqrt{5}\right)} & \frac{1}{2} \sqrt{\frac{7}{15}
   \left(3-\sqrt{5}\right)} & \sqrt{\frac{2}{3}} & \frac{\sqrt{2}}{3} \\
\end{array}
\right) \, , \label{matB1}
\eeq
while at width $l_{12} \eqsp 2$
\beq
\left(
\begin{array}{ccccc}
 \frac{3}{\sqrt{2}} & \frac{1}{2} \sqrt{\frac{3}{5} \left(23+3 \sqrt{5}\right)} & \frac{1}{2}
   \sqrt{\frac{3}{5} \left(23-3 \sqrt{5}\right)} & 2 & \frac{4}{\sqrt{3}} \\[1 mm]
 \frac{1}{2} \sqrt{\frac{3}{5} \left(23+3 \sqrt{5}\right)} & \frac{1}{5} \sqrt{2 \left(43+9 \sqrt{5}\right)} &
   \frac{6}{5} & \sqrt{3+\sqrt{5}} & \sqrt{\frac{10}{3}} \\[1 mm]
 \frac{1}{2} \sqrt{\frac{3}{5} \left(23-3 \sqrt{5}\right)} & \frac{6}{5} & \frac{1}{5} \sqrt{2 \left(43-9
   \sqrt{5}\right)} & \sqrt{3-\sqrt{5}} & \sqrt{\frac{10}{3}} \\[1 mm]
 2 & \sqrt{3+\sqrt{5}} & \sqrt{3-\sqrt{5}} & \sqrt{6} & \sqrt{2} \\[1 mm]
 \frac{4}{\sqrt{3}} & \sqrt{\frac{10}{3}} & \sqrt{\frac{10}{3}} & \sqrt{2} & \sqrt{6} \\
\end{array}
\right) \, . \label{matB2}
\eeq
Since the actual number values are not particularly informative we stop here. 

Naively, one could expect that the supersymmetry transformations $Q_1, \bar Q^4$ will relate these results to the matrices \eqref{matA2} and \eqref{matA3} of the previous section, up to some operator-specific factors. Due to the length changing of \eqref{susyChain1}, \eqref{susyChain2} at points 1 and 2, the $l_{12}$ matrix here ought to be related to the $l_{12}+1$ case there. Clipping off the last two rows and columns of the last two matrices and the first row and column of those of the last section we can form the (component-wise) ratios \eqref{matA2}/\eqref{matB1} and \eqref{matA3}/\eqref{matB2}. If the working hypothesis was right, the two ratios should be due to operator rescalings and therefore come out equal.
Somewhat disappointingly, this is not case. What went wrong? The renormalised theory has $Q,\bar Q$ and $S, \bar S$ (conformal) supersymmetry generators. The states we consider belong to long multiplets in the interacting theory, whose primary states are of the form $\tr[(X \bar X + Y \bar Y + Z \bar Z)Z^{L-2}]$. In an on-shell superspace correlator for the entire multiplets there are $\theta_I^\alpha, \bar \theta^{I \dot \alpha}, \, I \, \in \, \{1 \ldots 4\}$ Grassmann variables at each long point and half as many at the half-BPS point 3. Using all the supersymmetries we can remove as many Grassmann variables as there are in the long superfields. It follows that there must be two left and two right supersymmetry-invariant combinations of Grassmann variables which we shall call $\Lambda_{1,2}^\alpha, \bar \Lambda_{1,2}^{\dot \alpha}$. From these one can build \emph{nilpotent invariants}, cf.\ \cite{nilpotent}. In particular, there are (schematically) the nine combinations $(\Lambda_i \Lambda_j ) (\bar \Lambda_i \bar \Lambda_j)$ whose lowest order is $O(\theta^2 \bar \theta^2)$. The $\cC_{B, \bar B, D}$ correlators are hidden exactly at that order of the Grassmann expansion of the superspace three-point function, whereas the $\cC_{A, \check A, D}$ correlators are $O(\theta^3 \bar \theta^3)$ components. In conclusion, in these orders the superspace three-point function has 10 free coefficients multiplying the \emph{body} and the nine nilpotent invariants. The Grassmann expansion of the 10 different parts can result in correlator pairs that need not be proportional. The problem of fixing these coefficients is sizeable, but --- in the cases we discussed --- probably not hopeless at least in the leading orders in the coupling. From a super-function point of view this is quite an interesting future project! This would also shed light on the question whether the hexagon formalism can correctly handle all descendants under its non-manifest symmetries. Understanding the structure of this correlator on \emph{analytic superspace} would also be fascinating \cite{paulPaul}.

\subsection{Correlators of the form $\cC_{C, \tilde E, \tilde F}$} \label{cutCEF}

The lowest primary operator of the type $\cO_C^L$ with magnons $\{X,X,\Psi,\Psi\}$ occurs at length $L=6$. Solving the nested Bethe equations, we find the rapidities
\beq
u_j \, \in \, \left\{ \pm \frac{1}{2} \sqrt{1 + \frac{2}{\sqrt{5}}} \, , \ \pm \frac{1}{2} \sqrt{1 - \frac{2}{\sqrt{5}}} \, \right\} \, , \qquad v_k \, \in \, \left\{ \pm \sqrt{\frac{1}{2}} \, \right\} \, , \label{rootsL6G10}
\eeq
where $u_j$ are the four level-1 rapidities with $j=1,\dots4$ and the two auxiliary rapidities are $v_k$ with $k=1,2$. The energy, or one-loop anomalous dimension, is $E=10$ in all these cases. We restrict our attention to the case $\cC_{C, \tilde E, \tilde F}^{L_1, L_2, L_3}$ with $L
_1=6$, but contract it onto a variety of $u(1|1)$-sector operators $\cO_{\tilde E}^{L_2}$ at point 2 and $su(2)$-sector operators $\cO_{\tilde F}^{L_3}$ at point 3. For this latter operator we focus on $L_3=4,5,6,7$ with energies $E=6,4,5\pm\sqrt{5},4\pm 2$, respectively.
The coincidence of the series of anomalous dimensions (and underlying level-1 rapidities) of $\cO_{\tilde F}$ with those of $\cO_B, \, \cO_A$ is dictated by supersymmetry as stated by \eqref{susyChain1}, \eqref{susyChain2}. 

The $psu(1,1|2)$ Bethe ans\"atze yield a number of states with two equal fermions as excitations, and are thus candidates for $\cO_{\tilde E}$. However, some have infinite auxiliary rapidities signalling that these are descendants of the $\cO_A$ states, this time by the supersymmetries inside the sector in question. These states are the same as in a direct $su(1|1)$ computation, in which the only type of excitation on a chain of vacua $Z$'s is a fermion $\bar\Psi$. The Bethe equations are then simply those of free fermions:
\beq
e^{i \, p_j \, L_2} \eqsp 1 \, , \qquad i \, \in \, \{1,2\} \, .
\eeq
Admissible solutions for $\cO_{\tilde E}$ are non-degenerate whereby we end up with the same energies as for $\cO_F$ in the four cases $L_3=4,5,6,7$ from above,
but for shifted lengths $L_3 \rar L_2 \eqsp L_3 - 1$.

Since the magnons are transversal, the edges must satisfy $l_{12} \, \geq \, 2 \, \leq \, l_{31}$ and thus these three-point functions are always connected. The $su(2)$ sector operators are the same as in \eqref{wavSU2}, \eqref{normWave} and will therefore lead to the entangled states \eqref{entSU2}. The definitions for the $su(1|1)$ operator are essentially the same with the scattering matrix $A_{jk}$ replaced by $-1$. In both cases we obtain four partitions. The complicated operator at point 1 yields 96 different partitions. Several of the six parts of the multi-component wave function yield the same partition, and we can again collect them into a single splitting factor. Of the total of 1536 partitions, only 216 lead to non-vanishing products of hexagon amplitudes. Yet, since there are up to eight magnons on each hexagon, each individual term consists of up to 28 scattering matrices. Additionally, the splitting factors bring in the auxiliary rapidities. Therefore, we gave up on analytic factorisation.

Since there are three excitation-carrying operators, we start with all partitions on the back hexagon. The hexagon amplitudes do not show real poles between the rapidities of the second and third operator because their magnons $\bar \Psi_2$ and $\bar X$ are not conjugate. On the other hand, there are real double and single poles between points 1,2 as well as 1,3. Our set of rapidities can run into these poles in correlators with $\cO_{\tilde E}^5$ or $\cO_{\tilde F}^6$. Due to our restriction to $L_1=6$, they cannot be inserted simultaneously because the sum of lengths must be even.

We might use a small shift $\delta$ in the rapidities of $\cO_{\tilde E}^5$ or $\cO_{\tilde F}^6$, respectively, to regulate the particle-creation poles and then expand the respective correlator in $\delta$. The poles appear as first and second order poles in $\delta$. We can make them disappear by introducing fermion signs as before: bringing both fermions of the $su(1,1)$ operator to the front hexagon we have to insert a minus sign (both fermions are of the same type), but the rule \eqref{sigGamma} adopts the opposite sign. Hence we cannot associate a fixed sign with the $+2 \gamma$ rotation of the fermion normalisation; the sign rather depends on whether we augment or descend by two mirror transformations according to how the partitions are formed. For non-degenerate rapidities, as well as for degenerate rapidities caused by inserting the operator $\cO^6_{\tilde F}$, the regularised hexagon amplitudes correctly reproduce free field-theory three-point functions. We give explicit results of this computation in the later Sec. \ref{twistedCEF}, in particular in equation \eqref{n4CEF}. The simple-minded regulator seems to irrecoverably fail when $\cO^5_{\tilde E}$ is put into the three-point function. In Section \ref{twistedCEF} we will return to this example using a deformation as a regulator.

\subsection{Correlators in the $su(3)$ sector}

As a further test of the hybrid formalism introduced above, let us consider the calculation of correlators in the scalar $su(3)$ sector, spanned by the vacuum field $Z$, a transversal excitation $X$ and a longitudinal excitation $Y$. Similar to before, we consider a vacuum built of fields $Z$, on which $n_1$ excitations $X$ and $Y$ are placed. In the nested Bethe ansatz, the scalars $X$ correspond to level-1 excitations, and $Y$ to level-2 excitations whose number we denote by $n_2$.\footnote{We can similarly use $\bar{X}$ as level-1 excitation with $\bar Y$ as level-2 excitation, as well as $\bar{X}$ as level-1 excitation and $Y$ as level-2 excitation.}
The operators will be of the form 
\begin{align}
&\cO^L_H =  \tr(\hat Z^{L-2} X \hat{Y} ) \, ,
&&\cO^L_{\tilde{H}} =  \tr(\hat Z^{L-2} \bar{X} \hat{\bar{Y}} ) \, ,
&&\cO^L_{\check{H}} =  \tr(\hat Z^{L-2} \bar{X} \hat{Y} ) \, , 
\end{align}
and we study the structure constants $\mathcal C_{H,\tilde{H},D}^{L_1,L_2,L_3}$ and $\mathcal C_{H,\check{H},D}^{L_1,L_2,L_3}$.
The Bethe equations in the $su(3)$ sector can be written explicitly as
\begin{equation}
\begin{aligned}
    1 &= e^{ip_j L}\prod_{k \neq j}^{n_1} \frac{u_{j}-u_{k}-i}{u_{j}-u_{k}+i} \prod_{k = 1}^{n_2} \frac{u_{j}-v_{k}+\frac{i}{2}}{u_{j}-v_{k}-\frac{i}{2}}\,, \\
    1 &= \prod_{k = 1}^{n_1} \frac{v_j-u_k+\frac{i}{2}}{v_j-u_k-\frac{i}{2}} \prod_{k \neq j}^{n_2} \frac{v_j-v_k-i}{v_j-v_k+i}\,.
\end{aligned}
\end{equation}
Again, the momentum-carrying roots are given by $u_j$ and the auxiliary roots by $v_j$. In addition, there is the cyclicity constraint $\prod_{j=1}^{n_1} e^{i p_j}=1$. 

We restrict to operators with two excitations, in particular we consider $n_1=2$ and $n_2=1$, and thus $l_{12}=2$. 
For $L=3,4,5$ we find the following solutions to the Bethe equations
\vskip 0.2 cm
\begin{center}
    \begin{tabular}{c|c}
         $L$ & $u_1$ \\
         \hline 
        $3$ & $\frac{1}{2\sqrt{3}}$ \\
        $4$ & $\frac{1}{2}$ \\
        $5$ & $\frac{1}{2}\sqrt{1 \pm \frac{2}{\sqrt{5}}}$
    \end{tabular}
\end{center}

Similar to the construction in Sec.\ \ref{sec:Hybrid}, we can write
the Bethe wave function for $su(3)$ states with two excitations is schematically given by (up to a normalisation)
\begin{equation}
\begin{aligned}
    \psi^{1,1}(L) = &\sum_{m_1<m_2} \left(g_{XY} e^{i p_1 m_1 + ip_2 m_2 } + (g_{YX} R_{12}+g_{XY} T_{12})e^{ip_2 m_1 + ip_1 m_2} \right) \ket{X^{m_1} Y^{m_2}} \\
    + &\sum_{m_1<m_2}  \left(g_{YX} e^{i p_1 m_1 + ip_2 m_2 } + (g_{XY}R_{12}+g_{YX}T_{12})e^{ip_2 m_1 + ip_1 m_2} \right) \ket{ Y^{m_1} X^{m_2}} \,.
\end{aligned}
\end{equation}
The transmission and reflection amplitudes are
\begin{align}
    & T_{12}=\frac{A-B}{2}=\frac{u_1-u_2}{u_1-u_2+i}\, , && R_{12}=\frac{A+B}{2}=\frac{-i}{u_1-u_2+i}\, ,
\end{align}
with $A$ and $B$ the S-matrix elements \cite{beisertSu22}. Matching onto the results of the nested Bethe ansatz, the result can up to normalisation be simplified to 
\begin{equation}
    \psi(L) = \sum_{m_1<m_2} \left( e^{i p_1 m_1 + ip_2 m_2 } - e^{ip_2 m_1 + ip_1 m_2} \right) \left(\ket{ X^{m_1}  Y^{m_2}} - \ket{ Y^{m_1} X^{m_2}} \right)\,.
\end{equation}
Similar to Sec.\ \ref{secEntangled} we now cut this state to obtain the splitting factors as
\begin{align}
   \psi(l_1+l_2)&= \psi(l_1,\{ X_1, Y_2 \}) \psi(l_2,\{\}) - \psi(l_1,\{ Y_1, X_2 \}) \psi(l_2,\{\})\nonumber \\
   &+ e^{i p_2 l_1 } \left( \psi(l_1, \{X_1\}) \psi(l_2,\{Y_2\})  - \psi(l_1,\{Y_1\}) \psi(l_2,\{X_2\}) \right) \nonumber\\
   &+e^{i p_1 l_1 } \left(R_{12}-T_{12} \right) \left(\psi(l_1, \{X_2\}) \psi(l_2,\{Y(u_1)\}) - \psi(l_1,\{Y_2\}) \psi(l_2,\{X_1\})\right) \nonumber\\
   &+ e^{i (p_1+p_2) l_1 } \left(\psi(l_1,\{\}) \psi(l_2,\{X_1, Y_2\}) - \psi(l_1,\{\}) \psi(l_2,\{Y_1, X_2\}) \right)\,.
\end{align}

Again, the normalisation of the states can be obtained by generalising the Gaudin norm $\mathcal{G}$ in \eqref{normWave} to the \textit{full Gaudin determinant} in the $su(3)$ sector.
Putting the ingredients together, we reproduce the non-vanishing results obtained from field theory  
\begin{align}
    &\cC^{3,3,2}_{H,\tilde{H},D}=-\sqrt{2}\, ,
    &&\cC^{4,3,3}_{H,\tilde{H},D}=-\sqrt{3}\, , 
    &&\cC^{5\pm,3,2}_{H,\check{H},D}=\sqrt{5 \mp \sqrt{5}}\, , \nonumber \\
    &\cC^{5\pm,3,4}_{H,\tilde{H},D}=-\sqrt{2\pm\frac{2}{\sqrt{5}}}\, ,
    &&\cC^{5\pm,4,3}_{H,\check{H},D}=\sqrt{3\mp\frac{6}{\sqrt{5}}} \, ,
    &&\cC^{5\pm,4,3}_{H,\tilde{H},D}=\mp\sqrt{3\pm \frac{6}{\sqrt{5}}}\, .\label{resultSU3}
\end{align}
Further, we find the a set of vanishing correlation functions, namely $\cC^{3,3,2}_{H,\check{H},D}$, $\cC^{4,3,3}_{H,\check{H},D}$, $\cC^{4,3,5}_{H,\tilde{H},D}$, $\cC^{4,3,5}_{H,\check{H},D}$, $\cC^{5\pm,3,2}_{H,\check{H},D}$, $\cC^{5\pm,3,4}_{H,\check{H},D}$.

\section{Marginal deformations and structure constants} \label{secMarginal}

The \emph{dilatation operator} \cite{niklasDila} of the \nFour theory can be elaborated from Feynman integrals; its eigenoperators have well-defined anomalous dimension. At the planar one-loop level it acts as a prescription
\beq
\chi^A \chi^B \rar {\cal A}(A,B;C,D) \, \chi^C \chi^D \label{defDila}
\eeq
on two neighbouring fields (or their Yang--Mills derivatives) in a composite operator; for single-trace operators it can be given the interpretation of a spin-chain Hamiltonian. The exactly marginal deformations of $\cN=4$ SYM theory that preserve $\cN=1$ supersymmetry were classified in \cite{beta}.
For the $\beta$-deformed theory these result in a decoration of the planar amplitudes ${\cal A}$ in \eqref{defDila} by the $U(1)$ charges of the incoming and outgoing particles under the Cartan generators of the $su(4)$ flavour symmetry of the theory \cite{Khoze:2005nd,christophDila}. The fields have the following charges:
\begin{eqnarray}
&& Y=\Phi^{14} \ : \ \{1,0,0\} \, , \quad X=\Phi^{24} \ : \ \{0,1,0\} \, , \quad Z=\Phi^{34} \ : \ \{0,0,1\} \, , \nonumber
\\[1 mm] && \Psi^1 \ : \ \frac{1}{2} \, \{1,-1,-1\} \, , \quad \Psi^2 \ : \ \frac{1}{2} \, \{-1,1,-1\} \, , \label{charges} \\
&& \Psi^3 \ : \ \frac{1}{2} \, \{-1,-1,1\} \, , \quad \Psi^4 \ : \ \frac{1}{2} \, \{1,1,1\}~. \nonumber
\end{eqnarray}
The conjugate fields have the opposite charges and the derivatives $D^{\alpha\dot{\alpha}}$ are not charged.
Let us denote the charges of a field $\chi^A$ by the vector $q(A)=(q_1(A),q_2(A),q_3(A))$.
For two particles of type $A,B$ we define the \emph{charge wedge} as \cite{Lunin:2005jy}
\beq
q(A) \wedge q(B) \eqsp \sum_{i,j = 1}^3 \gamma_{ij} \, q_i(A) \, q_j(B)
\eeq
for three \emph{real} parameters $\gamma_{ij} \eqsp - \gamma_{ji}$ in the $\gamma$-deformation \cite{Frolov:2005dj}. In the rule \eqref{defDila} we scale
\beq
{\cal A}(A,B;C,D) \rar {\cal A}(A,B;C,D) \, e^{i \, \beta \bigl[ q(A) \wedge q(B) - q(C) \wedge q(D) \bigr]} \, . \
\eeq
We will use $\beta$ as an order parameter for Taylor expansions, while the $\gamma_{ij}$ are assumed to be of order $O(\beta^0)$. The $\beta$-deformation often discussed in the literature is the cyclic $\gamma$-deformation, i.e. $\gamma_{12} \eqsp \gamma_{23} \eqsp \gamma_{31} \, = \, 1$.

The conformal eigenoperators of \nFour SYM theory are affected by the $\beta$-deformation. Yet, as has been observed in \cite{raduNiklas}, the integrable structure of the spectrum problem is not hampered, rather the various $S$-matrices of the nested Bethe ansatz and with them the Bethe--Yang equations pick up deformation factors. For us this implies that \eqref{eq:BetheSU112} are deformed to
\begin{eqnarray}
1 & = & e^{i \beta (d_1n_1 + d_2n_2 +d_3n_3)} \prod_{j =1}^{n_1} e^{i p_j} \, , \nonumber \\
1 & = & e^{i p_i L} \, e^{i  \beta (d_{1} L + d_{12} n_2 + d_{13} n_3)} \, \prod_{j \neq i, \, j = 1}^{n_1} S^{11}(u_i,u_j) \, \prod_{j = 1}^{n_2} \frac{1}{S^{21}(v_j,u_i)} \, \prod_{j=1}^{n_3} \frac{1}{S^{31}(w_j,u_i)} \, , \nonumber \\
1 & = & e^{i \beta (d_{2} L + d_{21} n_1 + d_{23} n_3 )} \prod_{j = 1}^{n_1} S^{21}(v_i,u_j) \, , \\
1 & = & e^{i \beta (d_{3} L + d_{31} n_1 + d_{32} n_2)} \prod_{j = 1}^{n_1} S^{31}(w_i,u_j) \, . \nonumber
\end{eqnarray}
Here we introduced the real coefficients $d_i$ and $d_{jk}$, the latter being anti-symmetric in their indices, which we will match later against field-theory results to relate them to the $\gamma$'s in the sectors we are interested in. 
These relations arise from deforming the nested Bethe state by sending
\beq
S^{21} \rar e^{i \beta d_{21}} S^{21} \, , \qquad S^{31} \rar e^{i \beta d_{31}} S^{31} \, .
\eeq
Moreover, all momentum factors $e^{i p_i}$ are decorated by a deformation $e^{i \beta d_{1}}$; in the nesting philosophy one could call the entire combination $S^{10}$, describing the shift of a level-1 excitation over a vacuum site $Z$. New, as compared to the \nFour ansatz, are the $S$ elements
\beq
S^{20} \eqsp e^{i \beta d_{2}} \, , \qquad S^{30} \eqsp e^{i \beta d_{3}}
\eeq
for moving a level-2 and -3 magnon over one site $Z$. 
Likewise, we need to introduce deformed $S$ elements $S^{23}, S^{32}$ for bringing a left magnon through a right one, and vice versa. For every pair of magnons $\{ \Psi_j,  \bar \Psi_k\}$, we introduce a phase
\beq
S^{23}_{jk} \eqsp \Biggl\{ \begin{array}{ccl} m_j \, < \, m_k & : & e^{i \beta t_{32}} \\ m_j \, = \, m_k & : & 1 \\ m_j \, > \, m_k & : &  e^{i \beta t_{23}} \end{array} \label{defS23}
\eeq
with yet undetermined parameter $t_{23}=-t_{32}$.\footnote{
Following what we did in other parts of the nested Bethe ansatz, we could have placed all left magnons to the left of the right magnons at the beginning of the scattering, without a phase factor. Hence we would tend to put $t_{32} \eqsp 0$ and only work with $t_{23}$. But the only consistent choice seems to be the symmetric definition $t_{32} \, = \, -t_{23}$.}

Matching Bethe states onto dilatation-operator eigenstates in the two $psu(1,1|2)$ sectors \eqref{secX} and \eqref{secBar}, we find $t_{23} \eqsp d_{23}/2$. 
The $S^{ii}$ elements of the nested Bethe ansatz do not receive any deformation factor because the charge wedge is trivial in those cases. Likewise, the creation amplitudes remain unaltered. 
By studying the Bethe states corresponding to $\cO_A \ldots \cO_F$ and their anomalous dimensions, we fix the various parameters $d$ of the ansatz. For the two $psu(1,1|2)$ sectors with vacuum $Z$ we find
\begin{equation}
\begin{aligned}
    \{X,\Psi^4, \bar \Psi_1\}  :\, \, &  d_{12} = -\gamma_{12} + \gamma_{13}\,,\ d_{13} = \gamma_{12} - \gamma_{13}\,,\ d_{1} = \phantom{-} 2 \gamma_{23} \,, \\ 
    &d_{2} = \gamma_{13} - \gamma_{23},\ d_{3} = -\gamma_{13} - \gamma_{23} \, , \\
    \{\bar X,\Psi^3, \bar \Psi_2\}  :\,\, & d_{12} = -\gamma_{12} - \gamma_{13}\,,\ d_{13} = \gamma_{12} + \gamma_{13},\ d_{1} = -2 \gamma_{23}\,, \\
    &d_{2} = -\gamma_{13} + \gamma_{23},\ d_{3} = \gamma_{13} + \gamma_{23} \, .
 \end{aligned}
\end{equation}
Observe that the parameters of the second sector are obtained from those of the first by reversing the sign on $\gamma_{13}, \gamma_{23}$ whereas $\gamma_{12}$ stays put. For reasons detailed below we have studied the $\cO_B$ type of operators only in a reduced deformation with $\gamma_{12} \eqsp 0 \eqsp \gamma_{23}$. Then
\beq
\{X,\Psi^4, \bar \Psi_1\} : \quad d_{23} \eqsp \gamma_{13} \, , \qquad  \{\bar X,\Psi^3, \bar \Psi_2\} : \quad d_{23} \eqsp -\gamma_{13} \, .
\eeq

We will perform our structure-constant computations in the deformed theory in a perturbative expansion in $\beta$.
Expanding the Bethe equations up to $O(\beta^6)$, one can determine $u_i,v_j,w_k$ up to the same order (we exclude infinite rapidities for now). For the states in question, the Bethe roots receive corrections in terms of $\gamma_{23}$ at $O(\beta^1)$ while the other two parameters enter at $O(\beta^2)$.

\subsection{Deformed correlators in the $psu(1,1|2)$ sector} \label{chapter4}

Up to now we lack a construction principle for hexagon amplitudes in the presence of marginal deformations. Given that the nested Bethe ansatz for the spectrum problem of the undeformed theory can be amended by dressing various $S$-matrices by deformation factors \cite{raduNiklas}, can we do the same with \nFour hexagon amplitudes? A hint at this possibility is the structure of the various coefficients in the multi-component wave functions \eqref{superClean}, \eqref{superCleanC}, \eqref{ansFB1}, which will all individually pick up homogeneous deformations due to the products of $S^{21}$ matrices in \eqref{whatF}. Further, equating the multi-component wave functions with their nested equivalents, one can determine deformation factors for the components of the $S$-matrix in the former. Therefore, one finds that all splitting amplitudes, at least for $\cC_{A, \tilde A, D}, \, \cC_{B, \tilde B, D}$ partitions, have global deformation factors. But this property is not manifest and when it is lost in sums over coefficients,
we will split by independent powers of deformations at the expense of dealing with a larger set of partitions. These considerations will be made more explicit below, when revisiting the structure constants $\cC_{A, \tilde A, D}$, $\cC_{B, \tilde B, D}$ and $\cC_{C, \tilde E, \tilde F}$ in the deformed theory. In doing so, we will assume that the \nFour hexagon amplitudes will only pick up global $\beta$-dependent phase factors, whose explicit form we fix by comparison with field-theory results at tree-level in both the undeformed and deformed theory.

In the undeformed \nFour theory it does not matter which fermion pair is chosen for the $A, \tilde A$ operators in the first class of correlators. However, the deformation changes with the charges of the fermions. Nevertheless, there is an obvious map between the two cases such that it suffices to study one example. Similarly, in the third class of correlators we restrict to one of four possible types (and again $L_1 \eqsp 6$).
Note, that the deformation can regulate the particle-creation poles in \nFour hexagon amplitudes of $\cC_{A, \tilde A, D}$, $\cC_{B, \tilde B, D}$ because the rapidities of the second $psu(1,1|2)$ sector are different from those of the first unless $\gamma_{23} \eqsp 0 \eqsp \gamma_{12}$ or $\gamma_{23} \eqsp 0 \eqsp \gamma_{12}$. On the other hand, the $\cC_{C, \tilde E, \tilde F}$ correlators connect three distinct types of operators so that they are not very susceptible; almost any kind of twist can act as a regulator.

\subsubsection{A closer look at $\cC_{A, \tilde A, D}$} \label{twistingAAD}

The partitions of magnons into the two hexagons for this simplest class in our set of correlators were listed in \eqref{hexAmpsAAD}. We attribute a global deformation factor $e^{i \beta r_j}, \, i \, \in \, \{1 \ldots 10\}$ (with $r_j\in\mathbb{R}$) to each of these ten (products of) hexagon form factors. As previously pointed out, the first two cases each occur with four different permutations of the momenta, but they should all appear with the same (momentum-independent) deformation factor. Nevertheless, we have to keep track of the global deformation of the corresponding four splitting amplitudes.
 
Here we use $\beta$ as a regulator for the hexagon's particle-creation poles.\footnote{A similar idea was applied in \cite{McLoughlin:2020siu}, where $\beta$ was used to regulate degeneracies in a $1/N$-expansion around the planar theory.} In the study \cite{doubleTorus} this regularisation required inserting deformation factors on certain edges of higher-point correlators. We will encounter this phenomenon when discussing $\cC_{C, \tilde E, \tilde F}$ below. Since there are only two non-trivial operators in this example, we can create enough freedom to accommodate for such effects introducing arbitrary real shift parameters $s$, allowing us to modify the parameters $d$ in the \emph{states} but \emph{not} in the \emph{Bethe-Yang equations} (and thus their roots), i.e.\ for $\cO_A$
\beq
e^{i \beta (d_{1} + s_{1})}, \qquad e^{i \beta (d_{2} + s_{2})}, \qquad e^{i \beta (d_{21} + s_{21})}, 
\eeq
and for $\cO_{\tilde A}$
\beq
e^{i \beta (d_{1} + \tilde{s}_{1})}, \qquad e^{i \beta (d_{3} + s_{3})}, \qquad e^{i \beta (d_{31} + s_{31})} \, .
\eeq
The $s$ parameters move the location of the deformation on the three-point function. The splitting amplitudes now manifest the new parameters in their global deformation factors.

The marginal deformations have the potential to destroy the usual twisted translation \eqref{eq:TwistedTranslation}, as the latter mixes operators with different R-charges.
In order to circumvent these issues, we restrict to three-point functions with $l_{12} \eqsp 2$ for a start: In these restricted cases there are no $\la \hat Z(a=0) \hat Z(a=1) \ra$ propagators, only lines connecting the two transversal excitations at these positions. All shifted vacuum sites in the operators at points 1 and 2 are now contracted with the shifted vacuum $\cO_D^{L_3}$ at point 3 ($a=\infty$), whose length is therefore fixed to $L_3 \eqsp L_1 + L_2 - 4$.  
Next, if only $Z$ is projected out of the twisted vacuum $\hat{Z}$ at point 2, we are dealing with a conformal eigenoperator also in the deformed theory. We are obliged to trust the effective propagator for fermions, however this seems a mild assumption because the fermions are transversal excitations (i.e.\ they do not mix with the vacuum).

For the evaluation of the correlation function $\cC_{A, \tilde A, D}$ using the hexagons approach, let us initially assume $\gamma_{23} \, \neq \, 0$ with no condition on $\gamma_{12}, \, \gamma_{13}$. The particle-creation singularities of the hexagon amplitudes are seen as first and second order poles in $\beta \gamma_{23}$. Our strategy is to expand in $\beta$ and demand that 
\begin{enumerate}
\item poles cancel,
\item the $\beta^0$ part of the deformed hexagon sum reproduces the undeformed amplitude, and
\item the deformed hexagon computation reproduces free field-theory results for the deformed Bethe states up to operator rescalings.
\end{enumerate}
The third condition implies constraints on the parameters $r_i, \, s_j, \, \gamma_{kl}$ upon eliminating operator normalisations, for which we only assume that they have a regular Taylor expansion of the form $1+O(\beta)$ (to reproduce the undeformed case for $\beta\rightarrow 0$). It will be wise to study the constraints separately for cases in the deformed version of matrix \eqref{matA2} with non-degenerate rapidities (off-diagonal entries) and degenerate rapidities (diagonal entries). All these equations are homogeneous in the total power of $r_i, \, s_j, \, \gamma_{kl}$ which increases with the order in $\beta$. Nonetheless, at higher orders it is possible to single out linear constraints forming linear combinations of the equations and factoring.  
All cases in which the deformation factor acts as a regulator require a similar analysis. A particularly interesting situation is $\gamma_{23} \eqsp 0$ and $\gamma_{12} \, \neq \, 0 \, \neq \, \gamma_{13}$, in which case degenerate rapidities differ only at $O(\beta^2)$ so that a single particle-creation pole becomes a double pole in $\beta$. While the details of the various scenarios widely differ, it should suffice to discuss the most generic one with $\gamma_{23} \, \neq \, 0$ as an illustration.

\paragraph{The $\beta^{-2}$ and $\beta^{-1}$ parts.}
The double poles cancel in the sum over partitions without imposing any constraints on our free variables. However, at $\beta^{-1}$ we find the conditions 
\beq
r_5 \eqsp 2 \, r_1 + 2 \, r_2 - r_3  - r_8 - r_{10} \, \qquad r_6 \eqsp 2 \, r_1 + 2 \, r_2 - r_4  - r_7 - r_9 \,
. \label{solLead}
\eeq

\paragraph{The $\beta^0$ part.}
Next, we demand that the $\beta^0$ terms reproduce the result from the undeformed theory. This ought to happen for both permutations of the level-1 rapidities. By construction the resulting equations are homogeneous and of order 2 in powers of $r_j, \gamma_{kl}$. Combining the $L_1 \eqsp L_2\eqsp 3$ and $L_1 \eqsp L_2\eqsp 4$ equations (the higher ones do not carry independent information) implies
\begin{eqnarray}
r_{1} & = & s_{21} + 2 \, s_{2} - s_{31} - 2 \, s_{3} + 2 \, \gamma_{12} - 4 \, \gamma_{23} + r_{2} \, , \nonumber \\  r_{9} & = & 2 \, s_{2} - 2 \, s_{3} - 4 \, \gamma_{23} + 2 \, r_{2} - r_{7} \, ,\\ 
r_{10} & = & 2 \, s_{2} - 2 \, s_{31} - 2 \, s_{3} + 2 \, \gamma_{12} + 2 \, \gamma_{13} - 4 \, \gamma_{23} + 2 \, r_{2} - r_{8}  ,\nonumber 
\end{eqnarray}
or
\beq
\gamma_{13} \eqsp 0 \, , \qquad r_9 \eqsp 2 \, s_{2} - 2 \, s_{3} - 4 \, \gamma_{23} + 2 \,r_2 - r_7 \, . \label{singOnTheWay}
\eeq
To proceed with the analysis in the second case is considerably harder, as setting $\gamma_{13}=0$ reduces the amount of information available in the lower orders of the $\beta$-expansion. We have checked that the result is the $\gamma_{13} \eqsp 0$ case of a solution found below and for which reason we renounce on presenting the detailed argument.

There are two more independent constraints involving the six shift parameters and the five parameters $r_2, \, r_3, \, r_4, \, r_7, \, r_8$. Both conditions are quadratic in all variables, though. We leave these aside for the moment and move to the constraints of the third type listed above.

\paragraph{Off-diagonal cases.}
Now let us first consider the twelve non-singular off-diagonal correlators. We compare to free field theory using operators normalised as in \eqref{normWave}, but include the entire $\beta$-dependence. The leading order is the same as in the undeformed case. At linear order in $\beta$, we can solve for the arbitrary operator rescalings, leaving one free parameter.
At $O(\beta^2)$ we can solve for seven out of eight contributions to the rescalings, and additionally there are constraints on the parameters $r,s,\gamma$. Remembering $\gamma_{23} \, \neq \, 0$, one of them is linear and solved by
\beq
r_3 \eqsp r_4 + r_7 - r_8~.
\eeq
Further, after imposing this constraint, another one factorises and is solved by
\beq
r_8 \eqsp -s_{31} + \gamma_{12} + \gamma_{13} + r_7 \qquad \vee \qquad r_8 \eqsp -s_{21} - 
   \gamma_{12} + \gamma_{13} + r_4 \, . \label{firstAlt}
\eeq
The remaining conditions are then equivalent to what we skipped already once before. We deviate to the four diagonal cases.

\paragraph{Diagonal cases.}
Imposing the constraints derived so far, the correlators on the diagonal are finite in $\beta$ and $O(\beta^0)$ contains constraints of the form $(r_i \, r_j + \ldots)/\gamma_{23}^2$, which are again equivalent to the second order conditions we left aside before. Yet, at $O(\beta)$, we find some cubic equations with which one can make progress: for either solution \eqref{firstAlt} we find the solutions
\beq
r_7 \eqsp -s_{21} + s_{31} - 2 \, \gamma_{12} + r_4 \, , \label{nice}
\eeq 
or alternatively
\beq
r_7 \eqsp -4 \, s_{1} + 4 \, \tilde{s}_{1} + s_{21} - s_{31} + 2 \, \gamma_{12} \mp \frac{3}{2} \gamma_{13} - 
  \frac{11}{2} \gamma_{23} + 2 \, r_2 -r_4 \label{silly}
\eeq
with the $\mp$ referring to the first/second solution in \eqref{firstAlt}. Substituting equations \eqref{firstAlt} and \eqref{nice} into the quadratic equations we had so far ignored, yields
\begin{eqnarray}
r_4 & = & -2 \, s_{1} + 2 \, \tilde{s}_{1} + s_{21} - s_{31} + 2 \, \gamma_{12} - 8 \, \gamma_{23} + r_2 \quad \vee \label{goFurther} \\  r_4 & = & -2 \, s_{1} + 2 \, \tilde{s}_{1} + s_{21} - s_{31} + 2 \, \gamma_{12} + 4 \, \gamma_{23} + r_2 \nonumber
\end{eqnarray}
where the logical \emph{or} is not correlated with that of \eqref{firstAlt}. If instead of \eqref{nice} we use \eqref{silly}, we find the same constraints \eqref{goFurther}  plus some additional constraints $\gamma_{13} \eqsp \pm c \, \gamma_{23}$ with $c \eqsp 7,9$, which we can thus safely ignore.

Now, the $O(\beta^2)$ part of the diagonal correlators vanishes if we use the first possibility in \eqref{goFurther}, and for the second we find
\beq
\gamma_{12} \eqsp -\gamma_{13} \eqsp \gamma_{23} \qquad \vee \qquad \gamma_{12} \eqsp  \gamma_{13} \eqsp \gamma_{23} 
\eeq
and so the solutions can only exist in the cyclic ($\beta$-deformation) or anti-cyclic $\gamma$-deformation, respectively. There are no more constraints anywhere in the system of equations at $O(\beta^3)$, while we find
\beq
\gamma_{12} \eqsp \gamma_{23}
\eeq 
(with no constraint on $\gamma_{13}$) at the fourth order for the first choice in \eqref{goFurther}. The $\gamma_{13} \eqsp 0$ branch \eqref{singOnTheWay} of the analysis yields a sub-class of this last solution, although the proof is different. Finally, we observe that exact agreement between hexagon computation and free field theory can be achieved up to $O(\beta^5)$ for the diagonal cases (where we lose an order due to the simple poles in $\beta$) and $O(\beta^6)$ for the off-diagonal ones assuming the operator scalings to be absent, if
\beq
r_2 \eqsp -2 \, s_{1} - 2 \, \tilde{s}_{1} - s_{21} - 2 \, s_{2} + \gamma_{23} \, . \label{whatShift}
\eeq

\paragraph{Results for $l_{12}=2$:}
\begin{itemize}
\item $\gamma_{12} \eqsp \gamma_{23}$ and $\gamma_{13}$ free: \\
The global deformation factor of the splitting amplitudes offsets the extra factors $e^{ i \beta r_j}$ up to a global phase $e^{ i \beta \gamma_{13}}$ beyond which there is no twist on the individual hexagon amplitudes \eqref{hexAmpsAAD}. Hence, up to a simple unphysical phase, we can use the undeformed \nFour hexagon amplitude with deformed rapidities to correctly reproduce the tree-level structure constants of the deformed operators.
\item $\gamma_{12} \eqsp -\gamma_{13} \eqsp \gamma_{23}$ ($\beta$-deformation): \\
In the same way as above, there is a global phase $e^{-i \beta \gamma_{23}}$. The hexagons in the list  \eqref{hexAmpsAAD} with two excitations on the $0 \gamma$ edge receive extra deformation factors $e^{12\,  i \beta \gamma_{23}}$, those with two excitations on the $2 \gamma$ edge the inverse. There are no factors for excitations on the $4 \gamma$ edges.
\item $\gamma_{12} \eqsp \gamma_{13} \eqsp \gamma_{23}$: \\
The global phase is $e^{i \beta \gamma_{23}}$, the same rule for extra deformation factors applies as in the last case.
\end{itemize}
The last two solutions are quite remarkable: due to the extra deformation factors on the hexagon amplitudes, the particle-creation poles do not factor out exactly. These solutions exist owing to an intricate interplay of the Bethe roots' $\beta$-dependence with the extra deformation on the pairs of hexagons.

We have solved our equations for the $l_{12} \eqsp 2$ problem for the $r_j$ and were forced to accept some constraints on the $\gamma_{ij}$ whereas the shift parameters remained arbitrary. Fascinatingly we can understand from equation \eqref{whatShift} that the shifts merely contribute to a global phase; the asymmetry between the parameters of $\cO_A, \cO_{\tilde A}$ in that equation arises because we chose $r_2$ as the last parameter to be determined.

\paragraph{Allowing for edge widths $l_{12} \, \geq \, 2$.}
We now investigate the fate of the three solutions derived above for edge width $l_{12} \, >\, 2$, when the mismatch between the twisted translation and the marginal deformation starts to matter. Nevertheless, we assume no change in the definition of the twisted translation and use again the \nFour hexagon amplitudes with global deformation factors. 
For the solution in the first bullet point (so essentially the undeformed hexagon computation) we find that forcing all correlators to agree with \nFour SYM at $O(\beta^0)$ simply constrains the shift parameters:
\beq
s_{2} - s_{3} \eqsp 2 \, \gamma_{23} \, , \qquad s_{1} - \tilde{s}_{1} \eqsp - 4 \, \gamma_{23} \quad \vee \quad s_{1} - \tilde{s}_{1} \eqsp 14 \, \gamma_{23} \, . \label{condShift} 
\eeq
Choosing the shift parameters such that all deformation factors are removed in the states is a solution to the level-2,3 condition in the last equation and the first alternative at level-1. The second alternative seems mysterious. 

For the $l_{12} \eqsp 2$ solutions not equivalent to \nFour (second and third bullet point) something similar happens: the relation $s_{2} - s_{3} \eqsp 2 \, \gamma_{23}$ persists but the level-1 condition changes to
\beq
s_{1} - \tilde{s}_{1} \eqsp 2 \, \gamma_{23} \quad \vee \quad s_{1} - \tilde{s}_{1} \eqsp -16 \, \gamma_{23}~. 
\eeq
In all three cases, agreement between hexagon computation and free field theory \emph{cannot} be obtained beyond the leading order in $\beta$ if $l_{12} \, > \, 2$. 

At the beginning of this Subsection \ref{twistingAAD} we imposed $\gamma_{23} \, \neq \, 0$. The singular case $\gamma_{13} \eqsp 0$ is covered, and the case $\gamma_{12} \eqsp 0$ is excluded by the results from above. Further scenarios to case out are
\begin{itemize}
\item $\gamma_{23} \eqsp 0, \, \gamma_{12} \, \neq \, 0 \, \neq \, \gamma_{13}$. We had briefly mentioned this above. Demanding the absence of poles we quickly head for the undeformed model; however to make field theory and hexagon computation agree we would have to impose $\gamma_{12} \eqsp 0$ in contradiction to the assumption.
\item Both, putting $\gamma_{23} \eqsp \gamma_{13} \eqsp 0$ with $\gamma_{12}$ free, and $\gamma_{23} \eqsp \gamma_{12}  \eqsp 0$ with $\gamma_{13}$ free, yields degenerate rapidities at points 1,2. They can be regularised by a cut-off $\delta$ as $u_j(a=1) \eqsp u_j(a=0) + \delta$, then expanding in $\beta$ and finally in $\delta$. 
When $\gamma_{12}$ is non-zero this becomes inconsistent, whereas we fall upon the undeformed hexagon computation when $\gamma_{13}$ is non-zero.
\end{itemize}
In summary, there is no solution beyond those we derived for $\gamma_{23} \, \neq \, 0$. 

Reversing the sign on $\gamma_{13}$ and $\gamma_{23}$ the results carry over to three-point functions with the other fermion pair, i.e.\ with excitations $\{X, \bar \Psi_1\}$ at point 1 and $\{\bar X, \hat \Psi^3\}$ at point 2. As a consequence, the $\beta$-deformation $\gamma_{12} \eqsp \gamma_{23} \eqsp \gamma_{31} \eqsp 1$ yields consistent structure constants $\cC_{A, \tilde A, D}$ for one fermion pair, but not for the other, which would require the anti-cyclic version of the deformation.

The obstruction to using the \nFour amplitude in the deformed theory at $l_{12} \, > \, 2$ is due to constraints proportional to $\gamma_{12} \eqsp \gamma_{23} $. Sending both of them to zero, we can proceed.
Eliminating the twist from the splitting amplitudes setting 
\beq
-s_{1}  \eqsp 2 \, \gamma_{23} \eqsp \tilde{s}_{1} \, , \qquad s_{2} \eqsp -\gamma_{13} + \gamma_{23} \, , \qquad 
s_{3} \eqsp -\gamma_{13} - \gamma_{23}
\eeq
and then taking the limit $\gamma_{23} \rar 0$, we obtain exact agreement between the \nFour hexagon computation with $\gamma_{13}$-deformed rapidities and free field-theory three-point functions for the corresponding eigenoperators up to a global phase
\beq
e^{- i \, \beta \, \gamma_{13} (l_{12} - 2)} \label{phi13}
\eeq
for all the existing connected correlators in our set. We have done all the comparisons with the effective vacuum propagator of the \nFour hexagon construction. 

It is very tempting to interpret the last equation by a modification of the co-moving vacuum --- yet, what would be compatible with \eqref{uniPhaseC} below?
Also, note that dropping the additional phase $e^{i \beta \gamma_{13}}$ of our first deformation (cf.\ the first bullet point of the enumeration after equation \eqref{whatShift}) the exponent in the last equation becomes 
$- i \, \beta \, \gamma_{13} (l_{12} - 1)$.

\subsubsection{Correlators of the form $\cC_{B, \tilde B, D}$}

Because of the larger number of partitions, we will not attempt a detailed analysis of this example along the lines of the last section. At any rate, it became apparent there that using deformed rapidities with undeformed \nFour hexagon amplitudes can correctly reproduce tree-level results in the deformed theory for special deformations. The one-parameter case in which only $\gamma_{13} \neq 0$ seemed most promising. 
Therefore, we have studied the $\cC_{B, \tilde B, D}$ correlators only in the case of the pure $\gamma_{13}$-deformation, starting directly with undeformed \nFour hexagon amplitudes for $l_{12} \, \in \, \{1 \ldots 5\}$. The result is simple to state: The one-parameter case bluntly works, hexagon computations and projected field theory agree without extra phases.

\subsubsection{The correlator $\cC_{C, \tilde E, \tilde F}$ revisited} \label{twistedCEF}

In Section \ref{cutCEF} we failed to compute all \nFour correlators of the type $\cC_{C, \tilde E, \tilde F}$ when using a cut-off prescription for the particle-creation poles in the cases with degenerate rapidities. Can we use a $\gamma$-deformation as a regulator instead? Using the insights gained from the previous correlators, we will try to achieve this goal with the pure $\gamma_{13}$-deformation.

To compute those correlators in which rapidities can be degenerate between the operators at points 1 and 2, let us start the partitions on the back hexagon for $\cO_C^6$ and on the front one for $\cO_{\tilde E}$, such that the magnons from both operators are brought over the connecting edge $l_{12}$. We find that the explicit treatment of the third operator is not important for the evaluation, for definiteness let us start its partition on the back. On the other hand, for those cases in which the pole occurs between points 1 and 3, we place $\cO_C^6$'s excitations on the front hexagon and $\cO_{\tilde F}$'s excitations on the back one, so that the magnons are shifted over the common edge $l_{31}$. Again the positioning of the excitations for the third operator is arbitrary, and for definiteness we choose to start the partition on the back hexagon. With this prescription, we can extract the undeformed results for all 16 non-vanishing three-point functions from the $O(\gamma_{13}^0)$ terms. The results for the four non-singular cases agree between the two schemes, all others are covered in one of them. The resulting table of structure constants is
\beq
\cC_{C \tilde E \tilde F} \, = \, - \left(
\begin{array}{cccccc}
 * & \frac{2}{\sqrt{5}} & * & * & * & * \\[1 mm]
 3 \sqrt{\frac{3}{10}} & * & \frac{\sqrt{5}-1}{2 \sqrt{2}} & \frac{1+\sqrt{5}}{2
   \sqrt{2}} & * & * \\[1 mm]
 * & \frac{1}{5} \sqrt{25-2 \sqrt{5}} & * & * & \frac{3}{10}
   \sqrt{\frac{3}{2} \left(5+\sqrt{5}\right)} & \frac{1}{10} \sqrt{\frac{1}{2}
   \left(5+\sqrt{5}\right)} \\[1 mm]
 * & \frac{1}{5} \sqrt{25+2 \sqrt{5}} & * & * & \frac{3}{10}
   \sqrt{\frac{3}{2} \left(5-\sqrt{5}\right)} & \frac{1}{10} \sqrt{\frac{1}{2}
   \left(5-\sqrt{5}\right)} \\[1 mm]
 \frac{1}{2}\sqrt{\frac{3}{5}} & * & \frac{2}{\sqrt{5}} & \frac{2}{\sqrt{5}} &
   * & * \\[1 mm]
 \frac{1}{2 \sqrt{15}} & * & \frac{1}{\sqrt{5}} & \frac{1}{\sqrt{5}} & *
   & * \\
\end{array}
\right)\label{n4CEF}
\eeq
in full agreement with free field-theory results. The rows of the matrix are tied to $\cO^{L_2}_{\tilde E}$ with lengths $L_2=3,4,5,6$, the columns to $\cO^{L_3}_{\tilde F}$ with lengths $L_3=4,5,6,7$; an asterisk again denotes a non-existent combination of operator lengths. Note, that there is no phase factor $\sqrt{A_{jk}}$ for the $su(1|1)$ operators, for which the entangled state is build with the $S$-matrix $S_{su(1|1)}=-1$. 

It turns out we can do much better! According to the experience collected in \cite{doubleTorus} (and in agreement with what we saw in this work), magnons of operators with degenerate sets of rapidities should cross the connecting edge  without seeing deformation factors. Importantly, moving the deformation with our shift parameters does mean to make it disappear: it rather becomes localised on the edges not solicited in forming the partitions. It was also noticed in \cite{doubleTorus} --- albeit in a simpler situation featuring only $su(2)$ sector operators --- that magnons crossing the other end of such edges carrying a deformation ought to be affected. Let us thus write an ansatz 
\beq
e^{ i \beta \gamma_{13} (m_1 \, L_3 + m_2)} \label{putM1}
\eeq
for every magnon of $\cO_{\tilde F}^{L_3}$ moving from one part of the partition to the other in the situation where the rapidities at point 1,2 can be degenerate, and
\beq
e^{ i \beta \gamma_{13} (m_3 \, L_2 + m_4)} \label{putM2}
\eeq
when a magnon of $\cO_{\tilde E}^{L_2}$ is changing hexagon in the other case. Surprisingly, requiring that \emph{all} regularised hexagon calculations correctly reproduce the \nFour results is possible and singles out
\beq
m_2 \eqsp 2 - 6 \, m_1 \, , \qquad m_4 \eqsp 1 - 5 \, m_3 \, . \label{condM1}
\eeq
Furthermore, equating the two sets of 16 correlators is possible beyond $O(\beta^0)$, too, and implies 
\beq
m_1 \eqsp 0 \, , \quad m_3 \eqsp 1 \, . \label{condM2}
\eeq
These conditions are similar in spirit to \eqref{condShift} in the discussion of $\cC_{A, \tilde A, D}$. 

We find that all three-point functions of the pure $\gamma_{13}$-deformed theory computed via the deformed hexagon approach disagree from tree-level field-theory results by a common real factor
\beq
1+\frac{35}{24} \, \beta^2 \gamma_{13}^2+\frac{581}{384} \, \beta^4
   \gamma_{13}^4+\frac{13177}{9216} \, \beta^6
   \gamma_{13}^6 + \ldots \label{failReal}
\eeq
and some universal phases
\beq
e^{ i \, \beta \, \gamma_{13} \left(\frac{7}{2} + L_2 - L_3\right)} \, , \qquad e^{ i \, \beta \, \gamma_{13} \left(\frac{7}{2} - L_2 + L_3\right)} \label{uniPhaseC}
\eeq
in the two schemes pertaining to the 1,2 degenerate situation and the 1,3 degenerate one, respectively. 
The real factor \eqref{failReal} is actually a normalisation deficit for $\cO_C^6$, which we normalised according to \eqref{normWave} and misses unit norm by this factor.

In conclusion, comparing the hexagon results to field-theory correlators with correctly normalised operators, we find a simple length-depending phase similar to what we have seen in \eqref{phi13} for the $\cC_{A, \tilde A, D}$ correlators. Recall that such a phase is absent in the $\cC_{B, \tilde B, D}$ three-point functions. It would take more experience to make a guess for the general form of such phases --- \eqref{uniPhaseC} depends on where a deformation is inserted. It seems unlikely that they could be traced to a simple re-definition of the twisted translation.

\subsection{The $\beta$-deformed $su(2)$ sectors} \label{secSu2}

In the deformed higher-rank considerations above, it became apparent that only certain types of deformations can be made consistent when assuming that the deformation only introduces simple phase factors into the hexagon amplitudes. We suspect that this failure for the general $\gamma$-deformed case is related to the twisted translation, which mixes fields of different R-charges and could thus be expected to pick up $\gamma_i$-dependent phases as well.
In particular we failed to find a suitable generalisation of the hexagon approach that incorporates the $\beta$-deformation correctly. In this section we set out to understanding how a suitable $\beta$-deformed hexagon approach could be achieved by studying the simplest sectors in which the deformation occurs: the $su(2)$ sector with transversal excitations $X$, as well as the $su(2)$ sector with longitudinal excitations $Y$ on a vacuum of $Z$'s.

In the $\beta$-deformed theory, the Bethe equations in the $su(2)$ sector can be directly derived from the $\beta$-deformed dilatation operator, see e.g.\ \cite{McLoughlin:2020siu}, and are given by
\begin{equation}
    e^{i p_j L} \prod_{k \neq j}^n A^{\beta}(p_j,p_k)=1\,,
\end{equation}
similar to the undeformed case in \eqref{betYa}, where we now have a deformed S-matrix element $A^{\beta}(p_j,p_k)$ given by
\begin{equation}
    A^{\beta}(p_j,p_k)= -\frac{e^{i(p_j+p_k)}e^{2i\beta}+e^{-2i\beta}-2e^{i p_k}}{e^{i(p_j+p_k)}e^{2i\beta}+e^{-2i\beta}-2e^{i p_j}}\,.
\end{equation}
In principle, $\beta$ here corresponds either to the deformation parameter $\gamma_{13}$ or to $\gamma_{23}$, depending on whether we study the $su(2)$ sector containing $X$ or $Y$. In the following we will simply use $\beta$ and the correct $\gamma_{i3}$ can be reintroduced by specifying the $su(2)$ sector considered.

Introducing shifted momenta via $\tilde{p}=p+2\beta$ removes the explicit $\beta$-dependence from the S-matrix, which then takes on the form of the undeformed S-matrix element $A(u_j,u_k)$ of \eqref{aMat}. This comes at the cost of an explicit $\beta$-dependence of the Bethe equations and cyclicity constraint, 
\begin{equation}
    e^{i (\tilde{p}_j -2\beta) L} \, \prod_{k \neq j}  \, \frac{u_j - u_k - i}{u_j - u_k + i} = 1 \,, \qquad \text{and} \qquad \, \prod_{j=1}^{n_1} \, e^{i( \tilde{p}_2- 2\beta)} = 1\,,
    \label{eq:Bethe}
\end{equation}
where we introduced rapidities $u_j$ via the relation $\Tilde{p}_j=2 \, \mathrm{arccot}(2u_j)$.

In the following we compute structure constants of operators with two magnons, taken from both the $\beta$-deformed transversal and longitudinal $su(2)$ sector. In particular, we study three-point functions in these sectors where two operators carry two excitations each, while the third operator is a vacuum operator $\mathcal{O}_D^L=\text{Tr}(\hat{Z}^L)$. For the excited operators we again employ operators $\mathcal{O}_F^L$ as defined in \eqref{oDEF}, but also with the transversal $X$ replaced by the longitudinal $Y$. In order to highlight this, we collectively denote both types of operators as $\mathcal{O}_G^L$. Furthermore, we also study correlators involving excited operators which correspond to vacuum descendants in the undeformed theory but acquire non-vanishing anomalous dimension in the deformed case and denote them as $\mathcal{O}_D^{\prime L}$. Again these operators can carry either type of excitation.
We do not only compute these correlators at tree-level, but also at one-loop order in the deformed hexagon approach (neglecting wrapping) and compare to the respective field-theory results.

\subsubsection{Tree-level structure constants}

In the following we compute structure constants of operators taken from both the $\beta$-deformed transversal and longitudinal $su(2)$ sector. In particular, we study three-point functions in these sectors where the operators at $a=0$ and $a=1$ carry two excitations each, while the operator at $x=\infty$ is a vacuum operator $\mathcal{O}_D^L=\text{Tr}(\hat{Z}^L)$. To be able to Wick-contract, we insert the conjugate magnons on the second operator.
We perform these computations in a small $\beta$-expansion. 
Solving the twisted cyclicity condition and Bethe equations \eqref{eq:Bethe} perturbatively and for small lengths $L$ yields the following rapidities $u_\pm$   
\vskip 0.2 cm
\begin{center}
\begin{tblr}{c|l}
        Operator & Rapidities $u_{\pm}$ \\
        \hline
        $\mathcal{O}_D^{\prime 2}$ & $\frac{1\pm i}{4}\frac{1}{\beta} + \frac{-2 \pm i}{6} \beta + \frac{-8 \pm 7 i}{90}\beta^3 + O(\beta^5) $ \\
        $\mathcal{O}_D^{\prime 3}$ & $\frac{2 \pm i\sqrt{2}}{6} \frac{1}{\beta} + \frac{-8 \pm 3i\sqrt{2}}{24} \beta + \left(-\frac{7}{45} \pm \frac{67i}{480\sqrt{2}} \right) \beta^3 + O(\beta^5)$ \\
        $\mathcal{O}_G^4$ & $\pm \frac{1}{2\sqrt{3}} - \frac{2\beta}{3} \pm \frac{8 \beta^2}{9 \sqrt{3}} - \frac{16 \beta^3}{27} \pm \frac{112 \beta^4}{81 \sqrt{3}} +O(\beta^5)$ \\
        $\mathcal{O}_D^{\prime 4}$ & $\frac{3 \pm i\sqrt{3}}{8} \frac{1}{\beta} + \frac{-3 \pm i\sqrt{3}}{9} \beta + \frac{4}{405}(-21 \pm 8\sqrt{3}i) \beta^3 + O(\beta^5)$ \\
        $\mathcal{O}_G^5$ & $\pm \frac{1}{2}-\beta \pm \frac{5 \beta^2}{4}-\frac{11 \beta^3}{6} \pm \frac{145 \beta^4}{48} + O(\beta^5)$ \\
        $\mathcal{O}_D^{\prime 5}$ & $\frac{2 \pm i}{5}\frac{1}{\beta} + \frac{-8 \pm 5i}{24}\beta + \left(-\frac{23}{90} \pm \frac{211 i}{1152} \right) \beta^3 + O(\beta^5)$
    \end{tblr}
\end{center}
Here the two signs $(\pm)$ correspond to the rapidities of the two magnons.

As mentioned earlier, in the undeformed theory the Bethe equations allow for so-called vacuum descendants, i.e.\ solutions to the Bethe equations with infinite rapidities. As the energy of such operators vanishes, they are still BPS even though there are magnons on top of the vacuum. In the $\beta$-deformed theory however, infinite roots are no longer a solution due to the presence of the phase $e^{-i \beta L}$ in the deformed Bethe equations. This deformation lifts the degeneracy of these operators with the vacuum and their roots become proportional to $\beta^{-1}$ at leading order in a perturbative expansion in $\beta$.
Taking the limit $\beta \rightarrow 0$ sends the rapidities back to infinity and the operators become protected again (their energy vanishes). These operators are indicated as $\mathcal{O}_D^{\prime L}$, due to their relation to the vacua.

Again, we restrict here to the calculation of three-point functions with $l_{12} \eqsp 2$. This is done in order to avoid propagators of the form $\la \hat Z(a=0) \hat Z(a=1) \ra$, as these might need to be altered for the deformed theory.

Let us now use the hexagon form factor to compute three-point functions. 
For transversal excitations $X$ we find that the correlators can be straightforwardly evaluated by using the shift factor $e^{i (\tilde{p}-2\beta) l_{12}}$ ($=e^{i p l_{12}}$) for the partitions. The corresponding field-theory calculation (both with and without the twisted translation) gives the following results
\vskip 0.2 cm
\begin{center}
    \begin{tblr}{c|l}
        Correlator & Tree-level structure constant \\
        \hline
        $\cC^{4,2,2}_{G,D^\prime,D}$ & $-\frac{2}{\sqrt{3}} + \frac{8}{9\sqrt{3}}\beta^2 + \frac{112}{81\sqrt{3}} \beta^4 + O(\beta^6)$ \\
        $\cC^{4,3,3}_{G,D^\prime,D}$ & $-1 + \frac{4}{9}\beta^2 + \frac{56}{81} \beta^4 + O(\beta^6)$ \\
        $\cC^{4,4,4}_{G,D^\prime,D}$ & $-\frac{2\sqrt{2}}{3} +\frac{4\sqrt{2}}{27}\beta^2 + \frac{28\sqrt{2}}{81} \beta^4 + O(\beta^6)$ \\
        $\cC^{4,5,5}_{G,D^\prime,D}$ & $-\sqrt{\frac{5}{6}} - \frac{\sqrt{5}}{18\sqrt{6}}\beta^2 + \frac{133 \sqrt{5}}{648 \sqrt{6}} \beta^4 + O(\beta^6)$ \\
        $\cC^{5,2,3}_{G,D^\prime,D}$ & $-\sqrt{3}+\frac{\sqrt{3}}{2}\beta^2 +\frac{23}{8\sqrt{3}}\beta^4 + O(\beta^6)$ \\
        $\cC^{5,3,4}_{G,D^\prime,D}$ & $-\sqrt{2} +\frac{1}{\sqrt{2}}\beta^2 + \frac{23}{12\sqrt{2}} \beta^4 +O(\beta^6)$ \\
        $\cC^{5,4,5}_{G,D^\prime,D}$ & $-\sqrt{\frac{5}{3}} +\frac{5\sqrt{5}}{18\sqrt{3}}\beta^2 + \frac{517 \sqrt{5}}{648\sqrt{3}} \beta^4 + O(\beta^6)$ \\
        \hline[dashed]
        $\cC^{4,4,4}_{G,G,D}$ & $\frac{2}{3} - \frac{16}{27}\beta^2 - \frac{64}{81}\beta^4 + O(\beta^6)$\\
         $\cC^{5,4,5}_{G,G,D}$ & $\sqrt{\frac{5}{6}} - \frac{17}{18}\sqrt{\frac{5}{6}} \beta^2 - \frac{925 \sqrt{5}}{648 \sqrt{6}} \beta^4 + O(\beta^6)$\\
         \hline[dashed]
         $\cC^{3,3,2}_{D^\prime,D^\prime,D}$ & $\sqrt{2} +O(\beta^6)$ \\
         $\cC^{4,3,3}_{D^\prime,D^\prime,D}$ & $\sqrt{2} +\frac{2\sqrt{2}}{9}\beta^2 + \frac{22\sqrt{2}}{81} \beta^4 + O(\beta^6)$ \\
        $\cC^{4,4,4}_{D^\prime,D^\prime,D}$ & $\frac{4}{3} + \frac{16}{27} \beta^2 + \frac{64}{81} \beta^4 + O(\beta^6)$ \\
        $\cC^{5,4,5}_{D^\prime,D^\prime,D}$ & $\sqrt{\frac{5}{3}} +\frac{13}{18}\sqrt{\frac{5}{3}} \beta^2 +\frac{707}{648}\sqrt{\frac{5}{3}} \beta^4+ O(\beta^6)$
    \end{tblr}
\end{center}
In order to reproduce these results in the hexagon approach, we need to use the same Bethe equations for the operators at point 1 and 2. Therefore it seems as though in the hexagon picture operators at point 2 are conjugate with respect to operators at point 1 (despite being built on the same vacuum $\hat Z$ in the field theory). This is not obvious and should be studied further in future work. Despite these conceptual difficulties, our prescription reproduces the correct results also in the longitudinal case, as well as at one-loop level, as elaborated in the following.

The normalisation $\mathcal{N}$ of the three-point functions is given analogously to the undeformed case, only with the exception that the rapidities entering in the Gaudin norm $\mathcal{G}$ and the $S$-matrix depend on $\beta$.
For instance, for the operators $\mathcal{O}_G^4$ and $\mathcal{O}_G^5$ we find the Gaudin norm
\begin{equation}
    \begin{aligned}
        \mathcal{G}_{4} &= 108- 384 \beta^2 +\frac{1216}{3}\beta^4 - \frac{7168}{45} \beta^6 + O(\beta^6) \,,\\
        \mathcal{G}_{5} &= 80 - 440 \beta^2 +\frac{2000}{3}\beta^4 - \frac{1586}{9} \beta^6+ O(\beta^6) \,.
    \end{aligned}
\end{equation}
The Gaudin norm for the descendant operators $\mathcal{O}_D^{\prime L}$ starts contributing at order $O(\beta^4)$, while the hexagons involving a vacuum descendant start contributing at order $O(\beta^2)$. Due to the $1/\sqrt{\mathcal{G}}$ in the normalisation factor \eqref{normWave} the final result has leading order $O(1)$. This makes the calculation of correlators to a given order $\beta^k$ harder, as one has to use the rapidities up to order $\beta^{k+2}$. 

For longitudinal excitations $Y$, we need to dress the partitions by additional deformation factors $e^{2i\beta (d_{\alpha}-d_{\bar{\alpha}})}$, with parameters $d_{\alpha}$ depending on the partition $\alpha$ and the flavour, to find agreement with field theory. Specifically, the splitting factor becomes
\begin{equation}
    \omega_l(\alpha,\bar{\alpha})= (-1)^{|\bar{\alpha}|}\, \prod_{\Tilde{u}_i \in \bar{\alpha}} e^{2i\beta(d_{\alpha}-d_{\bar{\alpha}})} \left( \frac{u_i + \frac{i}{2}}{u_i - \frac{i}{2}} \right)^{l} e^{-2i \beta l} \prod_{u_1 \in \bar{\alpha}\,, u_2 \in \alpha}\frac{u_1 - u_2 - i}{u_1 - u_2 + i}\,, \label{eq:WeightfactorY}
\end{equation}
and we can calculate structure constants with hexagons as usual by using 
\begin{equation}
    \cC^{L_1,L_2,L_3}_{\mathcal{X},\mathcal{X},D} = \mathcal{N} \sum_{\substack{\alpha_1 \cup \bar{\alpha}_1= \{u_1, u_2\}\\
    \alpha_2 \cup \bar{\alpha}_2= \{u_3, u_4\}}}  \omega_{l_{12}}(\alpha_1, \bar{\alpha}_1) \omega_{l_{23}}(\alpha_2, \bar{\alpha}_2) \braket{\mathfrak{h} |\alpha_1,\alpha_2,\{\}} \braket{\mathfrak{h} | \bar{\alpha}_1,\{\},\bar{\alpha}_2}\,,
\end{equation}
where $\mathcal{X}$ represents any operator from the table above in one of the $su(2)$ sectors. 
We find agreement with field theory when using
\begin{align}
 & d_{\{\}}=0\, , &&d_{\{Y\}}=1\, , && d_{\{Y,Y\}}=2\, , &&d_{\{\bar{Y}\}}=-1~.
\end{align}
Further, we can easily include the transversal $su(2)$ sector by setting 
\begin{align}
    d_{\{X\}}=-d_{\{\bar{X}\}}~.
\end{align}
The resulting structure constants are the same as in the transversal case above.

\subsubsection{Asymptotic one-loop corrections}

In the following we test the proposed formalism to one-loop order in an expansion in the gauge coupling $g^2$.
For the hexagon computation two ingredients are necessary: the asymptotic hexagon and the contribution of virtual particles placed on the mirror-edges of the hexagon.
At one-loop order the virtual particles can only contribute if they are placed on an edge with width $l=0$. This is due to the fact that, in general, the contributions of virtual particles are suppressed by the bridge length $l$ and start at order $g^{2l +2}$. 
In the weak coupling limit, we can use $x^{\pm}(\Tilde{p}_j)=\frac{u_j \pm \frac{i}{2}}{g}$ to leading order in the coupling $g$. Solving the asymptotic Bethe equations at first loop order, we find one-loop corrections to the tree-level rapidities from above that are of the form
\vskip 0.2 cm
\begin{center}
    \begin{tblr}{c|l}
        Operator & One-loop corrections to rapidities $u_{\pm}$ \\
        \hline
        $\mathcal{O}_D^{\prime 2}$ & $\mp 4i \beta \pm \frac{8i}{3} \beta^3 + O(\beta^5)$ \\
        $\mathcal{O}_D^{\prime 3}$ & $\left( 2 \mp \frac{3i}{\sqrt{2}} \right) \beta - \left(-\frac{4}{3} \pm \frac{17i}{4 \sqrt{2}}\right) \beta^3 + O(\beta^5)$ \\
        $\mathcal{O}_G^4$ & $\pm \frac{4}{\sqrt{3}} - \frac{8}{3} \beta \mp \frac{16}{3\sqrt{3}}\beta^2 + \frac{64}{27} \beta^3 \pm \frac{64}{81 \sqrt{3}} \beta^4 + O(\beta^5)$ \\
        $\mathcal{O}_D^{\prime 4}$ & $\frac{8}{9}\left(3 \mp i \sqrt{3}\right) \beta - \frac{64}{81}(3 \mp 2i \sqrt{3}) \beta^3 + O(\beta^5) $\\
        $\mathcal{O}_G^5$ & $\pm \frac{5}{2} - \beta \mp\frac{15}{4}\beta^2 + \frac{5}{3} \beta^3 \pm \frac{15}{16}\beta^4 +O(\beta^5)$ \\
        $\mathcal{O}_D^{\prime 5}$ & $\left(3 \mp \frac{5i}{4}\right) \beta - \left(3 \mp \frac{245i}{96} \right) \beta^3 +O(\beta^5)$
    \end{tblr}
\end{center}

Finally, we need to include the measure factors $\mu(u_k)$ given in \cite{BKV} for the hexagon calculation.
Using the splitting factor from \eqref{eq:WeightfactorY}, we find the results for the one-loop corrections to structure constants as listed in the table below. 

In order to check the results of the hexagon approach with a field-theory computation, we make use of the results of \cite{Alday:2005nd} and \cite{David:2013oha} for undeformed and $\beta$-deformed $\mathcal{N}=4$ SYM theory, respectively. In these papers the authors realise that one-loop corrections of structure constants of scalar operators can be easily obtained from a Hamiltonian insertion into the respective three-point function. Using these methods, we find that the hexagon results are in agreement with the field-theory predictions.

\vskip 0.2 cm
\begin{center}
    \begin{tblr}{c|l}
        Correlator & One-loop corrections to structure constants \\
        \hline
        $\cC^{4,3,3}_{G,D^\prime,D}$ & $6 - \frac{68}{9} \beta^2 + O(\beta^4)$ \\
        $\cC^{4,4,4}_{G,D^\prime,D}$ & $4\sqrt{2} - \frac{56\sqrt{2}}{27} \beta^2 + O(\beta^4)$ \\
        $\cC^{4,5,5}_{G,D^\prime,D}$ & $\sqrt{30} - \frac{1}{18}\sqrt{\frac{5}{6}} \beta^2 + O(\beta^4)$ \\
        $\cC^{5,3,4}_{G,D^\prime,D}$ &  $4\sqrt{2} - \frac{5}{\sqrt{2}} \beta^2 + O(\beta^4)$  \\
        $\cC^{5,4,5}_{G,D^\prime,D}$ & $4\sqrt{\frac{5}{3}} + \frac{\sqrt{15}}{2} \beta^2 + O(\beta^4) $ \\
        \hline[dashed]
        $\cC^{4,4,4}_{G,G,D}$ & $-8 +  \frac{512}{27}\beta^2 + O(\beta^6)$\\
         $\cC^{5,4,5}_{G,G,D}$ & $-10 \sqrt{\frac{5}{6}} +  \frac{137}{6}\sqrt{\frac{5}{6}}\beta^2 + \frac{196}{81}\sqrt{\frac{5}{6}}\beta^4 + O(\beta^6)$ \\
    \end{tblr}
\end{center}

For structure constants involving operators $\cO_D^\prime$, one needs to use rapidities to higher orders in $\beta$ as they begin at order $\beta^{-1}$ which is computationally more expensive. Therefore we only include correlators of the form $\mathcal{C}_{G,G,D}$ and $\mathcal{C}_{G,D^\prime,D}$, and leave $\mathcal{C}_{D^\prime,D^\prime,D}$ for future work, though we do not expect any conceptual difference in this case.
We moreover disregard the extremal correlation functions (with $L_1=L_2+L_3$) at one-loop order. Here the bridge length $l_{23}$ vanishes and wrapping corrections contribute already at one-loop order. It would be interesting to understand how wrapping effects can be systematically included into the hexagon formalism in the deformed theory. This step also remains an obstacle in the undeformed case, and thus we only study the asymptotic hexagon in our work.

\section{Conclusions}

Integrability is a powerful tool, and provides methods to fully solve the spectrum problem of planar \nFour SYM theory via the (nested) Bethe ansatz. Together with conformal symmetry, this fully fixes the two-point functions of the model. 
For three-point functions the only missing ingredients are the structure constants, and the hexagon formalism \cite{BKV} allows to obtain these using integrability. 
This formalism allows for a straightforward application to rank-one sectors, whereas an extension to higher-rank cases is less direct due to the more complicated scattering processes whenever various types of flavours are present.
Therefore, most of the hexagon results in the literature are for operators in rank-one sectors. Our first and foremost achievement in this article is to have imported the necessary minimum of information from the nested Bethe ansatz into the hexagon construction to be able to work with operators from higher-rank sectors at all three points. 

The \emph{nested hexagon} of \cite{nestedHexagon} uses a similar approach, though trying to be more systematic. Indeed formula (19) there bears a resemblance to our expressions, although it does not contain e.g.\ the \emph{creation amplitudes} for higher-level magnons of the nested Bethe ansatz. This construction heavily relies on the construction of Bethe wave functions. We try to avoid this for a number of reasons: first, already in the spectrum problem the wave functions can have a complicated local structure. As an example we point to the $\cO_B$ operator we construct in Section \ref{secOB}; note in particular the complications due to the quadruple excitation at one site of the chain. 
Second, a characteristic (and often cumbersome) property of the hexagon operator is to bring out poles between conjugate particles at different edges due to crossing. It is hard to see how to recover the correct crossing properties from a wave function of the nested Bethe ansatz because only the $A$ element of the $S$-matrix \cite{beisertSu22} occurs, which does not yield the right pole structure under crossing. Yet, that would be necessary when considering three-point functions with more than one higher-rank operator. 
Third, it has been suggested to employ the hexagon as a tile in tessellations of $n$-point functions \cite{cushions,shotaThiago}. The prospect of constructing a single wave function for such an object seems quite daunting.

In our hybrid picture we only need the nested Bethe wave function to fix coefficients in the multi-component Bethe wave function, but afterwards employ the usual matrix hexagon formalism. These coefficients relay the information supplied by the auxiliary Bethe rapidities and, in the examples we considered, they have a very systematic structure: there is a tensor product of a left and a right factor, which should both be anti-symmetrised in the auxiliary rapidities. 
Our hybrid approach preserves the prettiest feature of the original hexagon construction: local details of the wave functions are totally eclipsed, all relevant information is in Bethe rapidities.
Despite these differences in the methods, it should be possible to compare results by considering three-point functions with only one higher-rank operator.
Note further that our three examples do not require wing symmetry, which is a condition in \cite{nestedHexagon}.

In this work we have only solicited two layers of the Bethe ansatz; one aim of future work will be to give concise formulae for the full ensemble. As suggested by the examples, the coefficients presumably have a determinant structure. On the level of the resulting wave functions \cite{nestedHexagon} reports on \emph{Slater determinants}.

In the second part of this work we raised the question, whether one can generalise the hexagon formalism of \nFour SYM to the $\gamma$-deformed theory.
It was observed in \cite{raduNiklas} that marginal deformations can be incorporated into the nested Bethe ansatz for the spectrum problem introducing deformation factors. 
However, since the defining axioms in the undeformed theory cannot easily be adapted to the deformed case, we tried to provide undeformed hexagons with deformation factors. The motivation for this strategy came from the study of sample correlators whose splitting amplitudes (or partition factors) pick up global deformation factors due to the deformation of the nested Bethe ansatz.

For a class of three-point functions of $psu(1,1|2)$ sector operators with a particularly small number of distinct hexagon amplitudes, we were able to exactly fix these factors. We found three solutions: the first is essentially the undeformed \nFour hexagon into which the deformation is only fed through the Bethe roots characterising the three operators; here a two-parameter deformation is possible. Remarkably, there are two additional special solutions with non-trivial deformation factors. These can only exist in one-parameter deformations akin to the $\beta$ one. However, in all three cases we have to choose the $R$-charges of the outer operators such that there are no effective vacuum-vacuum propagators $\la \hat Z \hat Z \ra$ between the two non-trivial operators on the three-point function. 

In the undeformed theory such propagators arise from the so-called twisted translation which is used to align the three operators at the standard positions $0,1,\infty$ along a line in Minkowski as well as flavour space. The underlying symmetry is broken by the deformations. While one would hope that the symmetry somehow survives, it is hard to see how to mend the twisted translation. For instance, a co-product structure as in \cite{Garus} does not map single-trace operators to their descendants since it does not respect cyclic symmetry.\footnote{We are grateful to T.\ McLoughlin, F.\ Seibold, and A.\ Sfondrini for many discussions on this point.}

Nevertheless, if the choice of operator lengths forbids the existence of $\la \hat Z_0 \hat Z_1 \ra$ propagators, we select eigenstates at both points with non-trivial operators even upon using the undeformed twisted translation, and thus the \nFour theory --- perhaps with decorated hexagon amplitudes as in our non-trivial examples --- can yield correct results. This is why  the $l_{12} \eqsp 2$ amplitudes in Section \ref{twistingAAD} are more prone to working out in the deformations. Surprisingly, in our example the situation improves if only the $\gamma_{13}$ parameter is non-vanishing. The complete \nFour hexagon amplitude then correctly reproduces the free field-theory three-point functions of the deformed operators independently of the lengths up to the phase \eqref{phi13}.

Due to the higher numbers of partitions the same analysis would be hard to adapt to our other examples of $psu(1,1|2)$ three-point functions. We limited our scope and directly tested whether the \nFour hexagon computation can match free field-theory structure constants of the deformed theory. For the second class of correlators there is an exact match, not even a running phase arises. For the third there is agreement up to a phase once again, barring for a common normalisation mismatch which we can trace to the normalisation of the operators. Here one would need to understand how the root of the full Gaudin determinant as an operator norm can be successfully superseded by some alternative definition in the deformed theory. However, in this class of three-point functions we have to re-insert deformation factors, a phenomenon that had previously been seen in tessellations of $su(2)$ sector correlators in \cite{doubleTorus}.

Problems with the twisted translation will be most pronounced when the longitudinal scalar excitations $Y, \bar Y$ are involved because these have off-diagonal propagators with the shifted vacuum. In all of the above we have therefore carefully excluded correlators involving longitudinal scalars. Only in Section \ref{secSu2} did we present some $su(2)$ sector examples in the $\beta$ deformation, which acts in the same way on operators involving only $Z, Y$, say, as the $\gamma_{13}$ deformation.

Note that the \emph{asymptotic} part of the coupling-constant dependence of all our correlators (so what is captured by the Zhukowsky variables $x^\pm(u)$ \cite{beiStau1,beiStau2}) should be easy to recover from the hexagon computations even in the presence of the deformations. 
We did so in the $\beta$-deformed $su(2)$ hexagon computation, and agreement with field theory is found: not only at tree level, but also for the one-loop contribution which we computed with methods of \cite{Alday:2005nd,David:2013oha}. Further, we only considered the asymptotic contribution at one-loop order. What happens to the \emph{gluing prescription} of \cite{BKV} is an interesting open question, though. Similar to the higher-rank correlators in the transversal $psu(1,1|2)$ sectors, we find that non-trivial deformation factors must be re-introduced through the splitting factor \eqref{eq:WeightfactorY} for longitudinal excitations. It is pressing to better understand these factors for three-point functions, cf.\ \eqref{putM1}, \eqref{putM2} and their consequences \eqref{condM1}, \eqref{condM2}, or in a similar vein \eqref{condShift}. This is a necessary step in the quest for a consistent hexagon approach in the $\gamma$-deformed theory. One might muse about the introduction of twist factors for the hexagon operator in a similar spirit as for the $S$-matrix in \cite{Ahn:2010ws}. It would further be interesting to make contact with the fishnet hexagon introduced in \cite{Basso:2018cvy} in a particular limit of the $\gamma$-deformed hexagon formalism.

\section*{Acknowledgements}

We are indebted to T.~McLoughlin, F.~Seibold, and A.~Sfondrini for many discussions about and common attempts at a systematic modification of the hexagon operator and the twisted translation in the presence of marginal deformations, as well as on spin-chain overlaps as an alternative route. We hope to build on the acquired experience in future work. B.~Eden is supported by Heisenberg funding of the Deutsche Forschungsgemeinschaft, grant Ed 78/7-1 or 441791296. D.~le~Plat is supported by the Stiftung der Deutschen Wirtschaft and would also like to thank KITP for the hospitality during the program \emph{Integrability in String, Field, and Condensed Matter Theory} as this research was supported in part by the National Science Foundation under Grant No. NSF PHY-1748958. A.~Spiering was supported by the research grant  00025445 from Villum Fonden and the ERC starting grant 757978, as well as by the European Union's Horizon 2020 research and innovation program under the Marie Sk\l odowska-Curie grant agreement No.\ 847523 `INTERACTIONS'.

\bibliography{refs}

\end{document}